\documentclass{emulateapj}

\newcommand{\1}{{~\sc i}}
\newcommand{\2}{{~\sc ii}}
\newcommand{\3}{{~\sc iii}}
\newcommand{\4}{{~\sc iv}}

\newcommand{\mic}{{\,$\mu$m}}

\shorttitle{Influence of the Environment on PAH Emission in Star-Forming Regions}
\shortauthors{Lebouteiller, Bernard-Salas, Whelan, Brandl, Galliano, Charmandaris, Madden, Kunth} 

\begin{document}


\title{Influence of the Environment on PAH Emission in Star-Forming Regions}


\author{V.\ Lebouteiller\altaffilmark{1,2}, J.\ Bernard-Salas\altaffilmark{1,3}, D.\ G.\ Whelan\altaffilmark{4}, B.\ Brandl\altaffilmark{5}, F.\ Galliano\altaffilmark{2}, V.\ Charmandaris\altaffilmark{6,7}, S.\ Madden\altaffilmark{2}, D.\ Kunth\altaffilmark{8}}
\email{vianney.lebouteiller@cea.fr}



\altaffiltext{1}{Center for Radiophysics and Space Research, Cornell University, Space Sciences Building, Ithaca, NY 14853-6801, USA}
\altaffiltext{2}{Laboratoire AIM, CEA/DSM-CNRS-Universit\'e Paris Diderot DAPNIA/Service d'Astrophysique B\^at. 709, CEA-Saclay F-91191 Gif-sur-Yvette C\'edex, France}
\altaffiltext{3}{IAS, B‰t. 121, Universit\'e Paris-Sud, 91435 Orsay, France}
\altaffiltext{4}{Department of Astronomy, University of Virginia, P.O. Box 400325, Charlottesville, VA 22904-4325, USA}
\altaffiltext{5}{Leiden Observatory, Leiden University, P.O. Box 9513, 2300 RA Leiden, The Netherlands}

\altaffiltext{6}{University of Crete, Department of Physics and Institute of Theoretical \& Computational Physics, GR-71003, Heraklion, Greece}
\altaffiltext{7}{IESL/Foundation for Research and Technology - Hellas,
 GR-71110, Heraklion, Greece, and Chercheur Associ\'e, Observatoire de
 Paris, F-75014, Paris, France}
\altaffiltext{8}{Institut d'Astrophysique de Paris (IAP), 98bis boulevard Arago, 75014 Paris, France}


\begin{abstract}
We investigate the emission properties of polycyclic aromatic hydrocarbons (PAHs) in various metallicity environments with the \textit{Infrared Spectrograph} on board \textit{Spitzer}. Local giant H\2\ regions are used as references as they enable access to the distinct interstellar medium components that contribute to the mid-infrared spectrum of star-forming galaxies: photodissociation regions (PDRs), photoionized gas, stellar clusters, and embedded regions. Three objects are considered, NGC\,3603 in the Milky Way, 30\,Doradus in the Large Magellanic Cloud, and N\,66 in the Small Magellanic Cloud. 
From the variations of the PAH/14\mic\ ratio, we find that PAHs are destroyed in the ionized gas for a radiation field such that [Ne\3]/[Ne\2]$\gtrsim3$. From the variations of the PAH/Hu$\alpha$ ratio, we find that the PAH emission sources in the giant H\2\ regions follow the same photodestruction law regardless of metallicity. 
We then compare these results with observations of starburst galaxies, H\2\ galaxies, and blue compact dwarf galaxies (BCDs).  While the integrated mid-infrared spectra of BCDs are reminiscent of a warm dusty ionized gas,  we observe a significant contribution to the PAH emission in starburst galaxies that is not arising from PDRs. 
\end{abstract}



\section{Introduction}\label{sec:intro}

From a simplistic chemical viewpoint, the abundance of molecules in the interstellar medium (ISM) of galaxies should scale with the metallicity of the environment, either because of the presence of nucleosynthetic constituents or because of the need for catalysts (dust grains). Obviously, indirect consequences of the low metal abundance also impact molecule survival, such as the lack of certain coolants (such as [O\1] or CO) that results in a globally warmer ISM, and hard interstellar radiation field (ISRF) due to energetic photons arising from low-metallicity ionizing stars. 

Observations show that nearby star-forming galaxies with metallicities lower than $\approx1/6$\,Z$_\odot$\footnote{We choose from now on the solar abundances of Asplund et al.\ (2006) for reference.} ($12+\log {\rm(O/H)}\lesssim7.9$) seem to lack CO molecules in giant molecular clouds (GMCs; Arnault et al.\ 1988; Taylor et al.\ 1998) and polycyclic aromatic hydrocarbons (PAHs) in photodissociation regions (PDRs) (Engelbracht et al.\ 2005; Madden et al.\ 2006; Wu et al.\ 2006; O'Halloran et al.\ 2006; Rosenberg et al.\ 2006). Even H$_2$, although it is detected in PDRs of metal-poor star-forming galaxies (Vanzi et al.\ 2000; Hunt et al.\ 2010), is not seen in absorption in their diffuse ISM (Vidal-Madjar et al.\ 2000; Lebouteiller et al.\ 2004; Thuan et al.\ 2002, 2005). This result illustrates the indirect effect of low-metallicity since the paucity of diffuse H$_2$ is  related to the combination of low H\1\ volumic density, dust grain abundance, and hard UV radiation (Vidal-Madjar et al.\ 2000).

In the present paper, we study the emission properties of PAHs, which are carbon-based molecules suggested by Leger \& Puget (1984) and Allamandolla et al.\ (1985) to explain the mid-infrared (MIR) emission bands seen ubiquitously toward star-forming regions. 
PAH features are often used as a diagnostic tool, as it is thought that PAH emission bands could trace star-formation on timescales on the order of O-stars lifetime (F{\"o}rster Schreiber et al.\ 2004; Weedman \& Houck 2008) or B-stars lifetime (Peeters et al.\ 2004).
It is therefore essential to understand the mechanisms of their formation and destruction as well as the physical conditions in which PAH emission occurs. 

The weakness of PAH emission in low-metallicity star-forming galaxies is thought to be due to a combination of several effects: insufficient chemical evolution of the environment leading to the lack of PAH progenitors (hydrocarbons), enhanced molecule destruction by energetic photons and by shocks, and  smaller volume of PAH-dominated regions. The main difficulty in investigating these parameters is their cross-correlation. Although a trend of decreasing PAH emission with both metallicity and ionizing radiation hardness has been found in several studies (Madden 2000; Madden et al.\ 2006; Wu et al.\ 2006; Engelbracht et al.\ 2006), the dominant parameter remains unknown. The study of H\2\ regions within M\,101 by Gordon et al.\ (2008) suggests that the aromatic band emission tends to depend more on the ionization of the environment rather than on metallicity. The photodestruction mechanism could be isolated by investigating PAH emission at ionization fronts in reflection nebul\ae\ (e.g.; Sellgren et al.\ 1985) and across spatially-resolved H\2\ regions where the metallicity is known to be uniform (Madden et al.\ 2006; Lebouteiller et al.\ 2007; Bernard-Salas et al.\ 2011). 

The present study aims at better understanding how the PAH emission in galaxies depends on the ISM morphology, by comparing the MIR properties of star-forming galaxies with those of giant H\2\ regions. The latter enable access to the distinct physical regions contributing to the composite mid-infrared (MIR) spectrum of star-forming galaxies, such as warm photoionized gas, PDRs, molecular clouds, shock fronts, stellar clusters, or supernov\ae\ remnants. Three well studied giant H\2\ regions with different metallicities are used as templates. The objects are NGC\,3603 in the Milky Way with essentially a solar metallicity, 30\,Doradus (hereafter 30\,Dor) in the Large Magellanic Cloud (LMC) with 0.6\,Z$_\odot$, and N\,66/NGC\,346 in the Small Magellanic Cloud (SMC) with 0.2\,Z$_\odot$. Although these regions might not be representative of typical H\2\ regions in star-forming galaxies, it is possible to distinguish their ISM components in the MIR to compare to integrated spectra of galaxies.

This paper is part of a series meant to investigate the MIR properties of nearby giant H\2\ regions. Lebouteiller et al.\ (2007) analyzed small spatial-scale variations of the MIR features in NGC\,3603 to characterize PAH destruction by energetic photons. A valuable advantage is that chemical abundances are remarkably uniform across the region, hypothesis confirmed a posteriori by Lebouteiller et al.\ (2008). Hence, it provided a unique possibility to isolate the metallicity parameter and study the specific effect of ionizing radiation hardness. The two other objects are being analyzed a similar way (Bernard-Salas et al.\ 2011; Whelan et al.\ 2011). 

We focus here on the brightest regions seen in the MIR to compare the PAH properties in the 3 giant H\2\ regions and to understand how PAH emission depends on both the physical and chemical conditions (Sect.\ \ref{sec:destr_ghiir}). We then view our results alongside integrated spectra of star-forming galaxies to determine general diagnostics of molecular and dust properties in various environments (Sect.\,\ref{sec:sfg}). The two specific issues we wish to address concern the influence of metallicity on the PAH-to-dust ratio, and whether the relative contribution of ionized gas versus PDRs in the integrated spectra of star-forming galaxies can explain the observed PAH deficiency in low-metallicity objects. Finally, we examine in Section \ref{sec:molh} the relation between PAH emission and molecular hydrogen.

\section{PAH emission properties}\label{sec:form_destr}

PAHs represent an important component in dust models (e.g.; Draine \& Li 2001) and their importance is multifold (e.g., Tielens et al.\ 2008; Galliano et al.\ 2008b). Most notably, they allow efficient cooling of the ISM, while providing most of the heating in PDRs (Bakes \& Tielens 1994; Hollenbach \& Tielens 1997). In addition, they contain a significant fraction ($\lesssim20$\%) of the carbon depleted from the gaseous phase (Allamandola, Tielens \& Barker 1989; Dwek et al.\ 1997; Zubko et al.\ 2004). We briefly review here the parameters responsible for the production and destruction of PAH molecules. We also emphasize the importance of the ISM morphology on the detection of PAH emission.

\subsection{Formation}\label{sec:form}

Carbon-rich asymptotic giant branch (AGB) stars provide a priori all the necessary ingredients (high collisional rates, cooling flow, shielding, periodic shocks) to produce significant amounts of PAH progenitors (Latter 1991). It is now thought that aliphatic hydrocarbons, which have been detected in large amounts in the envelopes of AGB stars (Cernicharo et al.\ 2001; Sloan et al.\ 2005; Boersma et al.\ 2006; Bernard-Salas et al.\ 2006), can lead to the formation of aromatic molecules via photoprocessing of hydrocarbon bonds. Theoretical studies support this PAH formation mode (e.g., Cadwell et al.\ 1994; Cherchneff 2006; Joblin et al.\ 2008).

With this scenario in mind, the PAH formation rate over a galaxy history could be mostly a function of the rate of low-mass stars entering the post-AGB phase. The spectral energy distribution modeling of galaxies by Galliano et al.\ (2008b) shows that PAH production in AGB stars is consistent with the PAH abundance evolution with metallicity (see also Mu{\~n}oz-Mateos et al. 2009). In particular, a young galaxy experiencing its first star-formation episode should contain little or no PAHs up until $\approx500$\,Myr. Based on this argument, Galliano et al.\ (2008b) proposed that delayed injection of PAHs by AGB stars is responsible for the low PAH abundance in metal-poor galaxies. If the dominant production process is indeed due to AGB stars, the PAH formation rate depends indirectly on the observed ISM metallicity, and the PAH abundance is expected to peak around solar metallicity sources (Galliano et al.\ 2008b), marking the transition from carbon-rich to oxygen-rich AGB stars (e.g., Renzini \& Voli 1981).

It must be noted that several other mechanisms have been proposed for PAH formation in the ISM, most notably splitting of graphitic grains in shocks (Tielens et al.\ 1987; Jones, Tielens \& Hollenbach 1996), accretion of C$^+$ in the diffuse ISM (Omont 1986, Puget \& Leger 1989), gas-formation by ion-molecule reaction in dense clouds (Herbst 1991), or photoprocessing of interstellar dust mantles (Greenberg et al.\ 2000). Most of these processes unfortunately still lack observational constraints.

\subsection{Destruction mechanisms}\label{sec:destr}

Molecule destruction eventually controls the PAH abundance in a galaxy, especially in extreme environments (Galliano et al.\ 2003; 2005; Madden et al.\ 2006). Metal-poor starburst galaxies globally exhibit harder radiation fields than more metal-rich galaxies since the radiation from ionizing stars have relatively lower opacity (e.g., Schaller et al.\ 1992; Schaerer \& Maeder 1992; Schaerer et al.\ 1993; Charbonnel et al.\ 1993).
As a result, the ISM is relatively warmer and the possible deficiency of some coolants (in particular CO) could exacerbate this effect. 
Furthermore, if massive stars in metal-rich starburst galaxies evolve in heavily extinguished regions (even opaque to MIR wavelengths), the observed interstellar radiation field (ISRF) would be found to be softer than in metal-poor galaxies (Rigby \& Rieke 2004). 

Radiation in low-metallicity environments is potentially energetic enough and intense enough for photons to destroy PAH molecules, depending on the molecule size (Omont 1986). PAHs can be heated to temperatures of $>1\,000$\,K by a single photon event. The energy deposited and thermalized into the molecule can lead to evaporation of bonds. Direct photolysis by UV photons is also an important process in the vicinity of young massive stars.

Observationally, PAH emission is seen closer to the ionizing source compared to H$_2$ and CO (e.g., Hollenbach \& Tielens 1999; Cesarsky et al.\ 1996). Photodestruction of PAHs effectively occurs in the ionized gas, where the observed intensity of PAH bands depends strongly on the physical conditions of the gas (e.g., Giard et al. 1994; Kassis et al. 2006). As opposed to H$_2$ and CO, PAH molecules are not self-shielded which, depending on the physical conditions, can lead to full inhibition of PAH emission at PDRs while H$_2$ and CO are still able to emit.

Finally, it must be added that sputtering can occur in strong shocks (e.g., Micelotta, Jones \& Tielens 2009) and in the hot gas, providing yet another means to destroy PAH molecules. However, shocks seem to be able to destroy all the dust content (Reach et al.\ 2002), so that they might not be good candidates to explain the observed deficiency of the PAH emission relative to dust continuum emission. In addition, rather than destroying PAHs, shocks could in fact result in molecule formation due to the splitting of large graphitic grains (Sect.\,\ref{sec:form}).

\subsection{PAH emission}\label{sec:em}

PAH emission arises in the neutral gas of galaxies, either diffuse (Haas et al.\ 2002) or dense (i.e., in PDRs, see introduction). Observations of integrated galaxies are more likely to show a significant contribution from PAHs in the diffuse ISM as compared to single star-forming region, where PAH emission mostly comes from PDRs, thus depending on the distribution of molecular clouds in the vicinity of ionizing stars. 

Both the PAH emission and the warm dust emission trace on first order the presence of young massive stars. However, while warm dust emits in the ionized gas of H\2\ regions (the grains in thermal equilibrium becomes hotter as the energy density $\mathcal{U}$ increases, Sect.\,\ref{sec:model}), PAH emission is intense at the vicinity of dense clouds such as GMCs. This different spatial origin can have important consequences in dwarf galaxies since these are prone to a lack of GMCs. In the SMC, Leroy et al.\ (2007) found that GMCs are either small or translucent with an average extinction several magnitudes lower than clouds in the Milky Way. Because of the low volume filling factor of GMCs combined with the beam size, the GMC properties in more distant objects is much less known. GMCs have been detected in nearby starburst galaxies and a few dwarf irregulars while non-detections are frequent in blue compact dwarf galaxies (BCDs) because of the low CO abundance in metal-poor environments that severely reduces the self-shielding of the CO core (e.g., Gondhalekar et al.\ 1998 and references therein). 

Although dwarfs contain H\1\ clumps with similar sizes and masses as the GMCs in the Milky Way (Lo et al.\ 1993), most of their H\1\ mass is expected to lie in the warm neutral medium. Indeed, in galaxies with a high star-formation rate, the low-density hot bubbles created by stellar clusters through stellar winds and successive SNe could well overlap so that the cold neutral medium is finally disrupted by the stellar feedback. PAH emission should thus be hampered since PAH molecules are destroyed in the warm ionized gas and in the hot gas (Sect.\,\ref{sec:destr}). 

As a conclusion, the PAH abundance, when compared to the dust component (warm or cold), strongly depends on the ISM morphology and in particular the distribution and number of molecular clouds. In peculiar objects such as BCDs, the observed lack of PAHs might thus not be a strong constraint on their actual formation/destruction over the galaxy history. Because of the dependence on the ISM morphology, only comprehensive galaxy chemical evolution models including GMC formation and stellar feedback would allow us to understand how the galaxy mass and the starburst properties might affect the integrated PAH emission in galaxies. In this study, we address this issue empirically by comparing the emission properties of single massive H\2\ regions with integrated star-forming galaxies.

\section{Data analysis}\label{sec:obs}

\subsection{Observations}\label{sec:obs_ghiir}

The observations of the giant H\2\ regions are presented in Lebouteiller et al.\ (2007; 2008) and we summarize here the most important aspects. NGC\,3603, 30\,Dor, and N\,66 were observed by the \textit{Infrared Spectrograph} (\textit{IRS}; Houck et al.\ 2004) onboard the \textit{Spitzer Space Telescope} (Werner et al.\ 2004) with the high- and low-resolution modules. This resulted in spectral coverage from 5\mic\ to 35\mic. Data were processed using the Spitzer Science Center S13.2 pipeline. We used the "basic calibrated data" product. Rogue pixels and on-the-fly flagged data were removed using IRSCLEAN\footnote{The IRSCLEAN software can be downloaded from \textit{http://ssc.spitzer.caltech.edu}}. Spectra were then extracted using the SMART\footnote{The SMART package can be found at \textit{http://isc.astro.cornell.edu/Smart/}} package (Higdon et al.\ 2004; Lebouteiller et al.\ 2010).

In contrast with Lebouteiller et al.\ (2007) and Bernard-Salas et al.\ (2011) who used the spatial information available along the low-resolution aperture, we consider in this study the spectra of the nominal positions for which both high- and low-resolution modules are available. The extraction aperture is $4.7''\times11.3''$ (full slit) for the \textit{Short-High} module, $11.1''\times22.3''$ (full slit) for \textit{Long-High}, $7.2''\times3.7''$ for \textit{Short-Low}, and $20.4''\times10.7''$ for \textit{Long-Low}. Spectra were normalized to the \textit{Short-High} spectrum based on matching the dust continuum and assuming a uniform light distribution within the apertures. Final spectra can be found in Lebouteiller et al.\ (2008). Since we consider only the nominal positions, there is an observational bias toward the MIR brightest sources within each giant H\2\ region. There are no dedicated observations of quiescent regions, except the positions N66\#1, \#2, and \#3 which are mispointings. 

The \textit{IRS} spectra of giant H\2\ regions probe various stellar/ISM components (Lebouteiller et al.\ 2007; Bernard-Salas et al.\ 2011), with most notably:\\
\textbf{Central star clusters.} A strong continuum has been detected shortward of 8\mic\ in NGC\,3603. The MIR spectrum is otherwise similar to that of the warm photoionized gas. \textit{Spitzer/IRAC} images show a dust-free cavity around the central clusters of NGC\,3603 and 30\,Dor, probably carved out by the winds of massive stars, with little or no MIR emission. Silicate dust emission was detected in emission toward, or close to, the central clusters of the 3 giant H\2\ regions, hinting at low optical depth lines of sight (Whelan et al. 2010). \\
\textbf{Warm photoionized gas.} It is detected via the H\1\ recombination line Hu$\alpha$ and the forbidden lines [S\3], [S\4], [Ne\2], [Ne\3], [Ar\3], and [Fe\3]. The ionized gas component harbors a strong and steeply increasing dust continuum longward of 10\mic.\\
\textbf{PDRs.} Each giant H\2\ region contains several individual PDRs. These are located further away than the ionized gas counterpart from the central cluster, indicating that the radiation coming from the central cluster dominates the energetics. PDRs show bright PAH bands as well as significant H$_2$, [Ar\2], [Si\2], and [Fe\2] line emission.\\
\textbf{Embedded MIR bright regions.} Embedded source show silicate dust in absorption. Only 30\,Dor and N\,66 show sources with deep silicate absorption in our observation. PAHs are also detected toward these regions in some cases.

Since we are dealing with extended sources, the extraction aperture (Sect.\,\ref{sec:obs_ghiir}) probes larger physical regions in more distant objects. Therefore, while observations of NGC\,3603 ($\approx7$\,kpc away, e.g., Moffat et al.\ 1983) can be centered on a single PDR, observations of 30\,Dor ($\approx49$\,kpc, e.g., Selman et al.\ 1999) and N\,66 ($\approx61$\,kpc, e.g., Hilditch et al.\ 2005) can include several PDRs. As an illustration, the projected size of the \textit{IRS} \textit{Short-High} aperture is 0.4\,pc in NGC\,3603, 2.7\,pc in 30\,Dor, and 3.3\,pc in N\,66. Furthermore, projection effects prevent us from asserting with accuracy the physical association between the tracers of the ionized gas and those of the PDRs. Nevertheless, observations were designed to study the brightest MIR knots which are dominated by a single region within the aperture with little contribution from nearby regions (see also Lebouteiller et al.\ 2007).

In the following we compare the giant H\2\ regions with a sample of BCDs (Wu et al.\ 2006), starburst galaxies (Brandl et al.\ 2006; Bernard-Salas et al.\ 2009), and H\2\ galaxies (Smith et al.\ 2007).

\subsection{Tracers}\label{sec:obsparam}

\subsubsection{Molecules and dust}\label{sec:obsparam_moldust}

We measure PAH emission by summing the flux of the most prominent bands in the \textit{IRS} wavelength range, i.e., 5.7\mic, 6.2\mic, 7.7\mic, 8.6\mic, and 11.3\mic. PAH features were measured using the PAHFIT profile decomposition algorithm (Smith et al.\ 2007). The 11.3\mic\ PAH flux was taken as the integrated flux of the 11.23\mic\ and 11.33\mic\ PAHFIT features. Similarly, the 7.7\mic\ PAH flux represents the integrated flux of the 7.60\mic\ and 7.85\mic\ PAHFIT features. The H\1\ recombination lines at 7.46\mic\ (Pf$\alpha$) and 7.50\mic\ (Hu$\beta$) contaminate the 7.7\mic\ PAH complex by up to 5\%. We added these lines manually in PAHFIT to obtain a satisfactory fit of the PAH complex. The plateau underneath the 7.60\mic\ and 7.85\mic\ features was included in the fit but ignored in the calculations of the PAH band fluxes, as it probably originates from larger molecules such as PAH clusters with several 100s carbon atoms (e.g., Bregman et al.\ 1989). PAH fluxes for the giant H\2\ regions are reported in Table\,\ref{tab:pahs}.

\begin{deluxetable*}{lllllll}
\tablewidth{0pc}
\tabletypesize{\scriptsize}
\tablecaption{PAH and continuum fluxes in giant H\2\ regions
.\label{tab:pahs}}
\tablehead{ \colhead{Position$^{\rm a}$} & \colhead{PAH 5.7\mic} & \colhead{PAH 6.2\mic} & \colhead{PAH 7.7\mic} & \colhead{PAH 8.6\mic} & \colhead{PAH 11.3\mic} & \colhead{$f$(14\mic)} }
\startdata
NGC3603\#1 &  0.22 (0.51) & 12.57 (0.91) & 20.18 (0.51) &8.6 (0.51) &11.16 (0.78) & 22.01 \\ 
NGC3603\#3 &  43.01 (0.83) & 301.09 (0.95) & 678.82 (0.83) &210.26 (0.83) &401.7 (3.5) & 799.55 \\ 
NGC3603\#4 &  51.53 (20.12) & 320.55 (15.62) & 500.42 (20.12) &159.3 (20.12) &345.09 (19.53) & 1332.09 \\ 
NGC3603\#5 &  52.71 (0.85) & 418.98 (0.95) & 947.76 (0.85) &289.61 (0.85) &323.87 (3.33) & 865.57 \\ 
NGC3603\#6 &  \nodata  & 15.32 (1.2) & \nodata  &11.88 (0.76) &14.39 (3.79) & 309.55 \\ 
NGC3603\#7 &  12.15 (1.94) & 35.89 (1.4) & 24.21 (1.94) &35.24 (1.94) &8 (0.68) & 8371.07 \\ 
\tableline
30Dor\#2 &  \nodata  & 10.03 (0.31) & 25.17 (0.34) &8.86 (0.79) &15.3 (0.47) & 10.27 \\ 
30Dor\#3 &  1.13 (0.63) & 4.15 (0.44) & 5.08 (0.63) &1.94 (0.63) &4.61 (0.6) & 88.02 \\ 
30Dor\#4 &  2.52 (0.47) & 3.62 (0.36) & 2.37 (0.47) &\nodata  &9.85 (0.22) & 230.33 \\ 
30Dor\#5 &  4.34 (0.51) & 38.11 (0.41) & 84.69 (0.51) &29.71 (0.51) &40.45 (0.76) & 33.74 \\ 
30Dor\#6 &  3.27 (0.77) & 40.56 (0.65) & 83.25 (0.77) &26.62 (0.77) &58.58 (0.92) & 156.98 \\ 
30Dor\#7 &  1.43 (0.59) & 16.87 (0.54) & 23.78 (0.59) &9.23 (0.59) &20.05 (0.65) & 308.08 \\ 
30Dor\#8 &  9.56 (0.64) & 67.53 (0.54) & 153.54 (0.64) &46.96 (0.64) &85.06 (0.86) & 224.46 \\ 
30Dor\#10 &  8.38 (0.58) & 63.08 (0.47) & 145.14 (0.58) &45.27 (0.58) &70.42 (0.75) & 133.5 \\ 
30Dor\#11 &  3.52 (0.68) & 26.44 (0.4) & 48 (0.68) &19.84 (0.68) &41.64 (0.79) & 168.71 \\ 
30Dor\#12 &  3.23 (0.62) & 30.26 (0.53) & 75.88 (0.62) &22.53 (0.62) &43.29 (0.88) & 140.84 \\ 
30Dor\#13 &  2.25 (0.56) & 13.17 (0.38) & 9.7 (0.56) &0.67 (0.56) &6.15 (0.37) & 38.14 \\ 
30Dor\#14 &  4.5 (0.6) & 48.79 (0.51) & 120.79 (0.6) &36.38 (0.6) &79.95 (1) & 467.99 \\ 
30Dor\#15 &  3.59 (0.63) & 58.87 (0.46) & 124.11 (0.63) &32.65 (0.63) &87.99 (0.99) & 635.24 \\ 
30Dor\#16 &  \nodata  & 16.39 (0.37) & 29.9 (0.43) &7.44 (0.19) &18.35 (0.82) & 466.53 \\ 
30Dor\#17 &  10.08 (1.11) & 30.04 (0.62) & 71.59 (1.11) &52.68 (1.11) &53.56 (1.02) & 176.05 \\ 
\tableline
N66\#1 &  \nodata  & 2.27 (0.12) & 4.55 (0.46) &0.33 (0.30) &3.39 (0.19) & 4.25 \\ 
N66\#2 &  0.18 (0.12) & 3.54 (0.09) & 1.33 (0.12) &1.59 (0.12) &5.48 (0.16) & 2.35 \\ 
N66\#3 &  0.53 (0.12) & 1.4 (0.09) & \nodata  &0.43 (0.12) &1.18 (0.1) & 1.61 \\ 
N66\#5 &  \nodata  & 2.88 (0.48) & 9.24 (0.89) &1.12 (0.42) &4.72 (0.62) & 39.02 \\ 
N66\#6 &  0.9 (0.55) & 3.62 (0.35) & \nodata  &4.05 (0.55) &4.27 (0.3) & 31.84 \\ 
N66\#7 &  \nodata  & 2.81 (0.29) & 4.15 (0.56) &0.75 (0.52) &3.1 (0.41) & 5.72 \\ 
N66\#8 &  0.03 (0.4) & 2.84 (0.36) & 5.26 (0.4) &1.18 (0.4) &4.12 (0.43) & 4.84 \\ 
N66\#9 &  \nodata  & 5.28 (0.34) & 9.94 (0.97) &2.63 (0.35) &7.93 (0.5) & 12.76 \\ 
N66\#10 &  0.08 (0.47) & 2.23 (0.32) & 5.22 (0.47) &0.68 (0.47) &2.77 (0.51) & 7.04 \\ 
N66\#11 &  0.46 (0.52) & 1.41 (0.31) & 0.64 (0.52) &0.52 (0.52) &2.28 (0.27) & 7.19 \\ 
N66\#12 &  0.5 (0.42) & 1.3 (0.31) & \nodata  &0.23 (0.42) &0.49 (0.23) & 5.43 \\ 
N66\#13 &  \nodata  & 2.36 (0.34) & 1.07 (0.25) &0.52 (0.31) &2.3 (0.52) & 3.08 \\ 
\enddata
\tablecomments{Fluxes are given in $\times10^{-20}$\,W\,cm$^{-2}$. The 14\mic\ continuum flux is integrated on a 1\mic\ wide window centered at 14.3\mic. The uncertainty is given in parentheses. We assume an uncertainty of 10\%\ for the 14\mic\ continuum due to the instrument flux calibration. PAH fluxes were measured using PAHFIT (Sect.\,\ref{sec:obsparam_moldust}).}
\tablenotetext{a}{Sources labels and coordinates are defined in Lebouteiller et al.\ (2008).}
\end{deluxetable*}

While Smith et al.\ (2007) measured PAH emission with PAHFIT in H\2\ galaxies, a spline method was originally used to remove the underlying spectral continuum in the BCDs and in the starbursts. Hence for consistency, we re-calculated PAH fluxes in these objects using PAHFIT (Table\,\ref{tab:sfg}). As expected, the new PAH fluxes are different than those measured by Wu et al.\ (2006), Brandl et al.\ (2006), and Bernard-Salas et al.\ (2009), while variations in band ratios are mostly unchanged. A discussion on the comparison between the spline method and the profile decomposition method can be found in Galliano et al.\ (2008a).

\begin{deluxetable*}{lllllllll}
\tablewidth{0pc}
\tabletypesize{\scriptsize}
\tablecaption{PAHFIT and continuum fluxes in star-forming galaxies
.\label{tab:sfg}}
\tablehead{ \colhead{} & \colhead{$Z_\textrm{MIR}$\tablenotemark{a}} & \colhead{$Z_\textrm{opt}$\tablenotemark{b}} &  \colhead{PAH 5.7\mic} & \colhead{PAH 6.2\mic} & \colhead{PAH 7.7\mic} & \colhead{PAH 8.6\mic} & \colhead{PAH 11.3\mic} & \colhead{$f$(14\mic)} }
\startdata
BCDs & & & & & & & & \\
\tableline
CG\,0752 &  \nodata & \nodata & 0.8 ( 0.21)  &4.36 (0.17) &14.06 (0.53) &2.1 (0.13) &3.53 (0.26) &6.39 \\
Haro\,11 &  0.56 & 0.18 & 5.11 ( 0.25)  &9.65 (0.12) &16.3 (0.49) &2.65 (0.38) &4.11 (0.17) &90.24 \\
IIZw40 &  0.31 & 0.28 & \nodata ( \nodata)  &1.94 (0.12) &3.61 (0.84) &1.05 (0.44) &2.05 (1.1) &79.17 \\
Mark\,1450 &  0.69 & 0.21 & 0.14 ( 0.04)  &0.3 (0.02) &0.34 (0.25) &0.19 (0.03) &0.19 (0.03) &1.99 \\
Mark\,1499 &  0.55 & 0.29 & 0.13 ( 0.03)  &0.78 (0.03) &1.91 (0.07) &0.58 (0.02) &0.82 (0.04) &0.84 \\
NGC\,1140 &  0.83 & 0.46 & 1.16 ( 0.22)  &6.41 (0.17) &15.42 (0.65) &4.28 (0.12) &7.18 (0.36) &8.65 \\
NGC\,1569 &  0.48 & 0.3 & 1.03 ( 0.26)  &8.77 (0.2) &17.77 (1.11) &4.94 (0.31) &10.9 (0.39) &78.45 \\
UGC\,4274 &  \nodata & \nodata & 0.67 ( 0.14)  &4.81 (0.09) &11.85 (0.77) &3.4 (0.08) &5.11 (0.27) &4.66 \\
UM\,461 &  0.19 & 0.13 & \nodata ( \nodata)  &\nodata  &\nodata  &\nodata  &\nodata  &1.74 \\
\tableline
Starbursts & & & & & & & & \\
\tableline
IC\,342 &  1.86 & 1.55 & 0 ( 1.3)  &264 (0.58) &569 (4.06) &189 (1.55) &303.4 (0.88) &343.78 \\
Mark\,52 &  \nodata & \nodata & 0.93 ( 0.3)  &14.3 (0.3) &35.5 (0.87) &9.07 (0.21) &12.46 (0.36) &34.23 \\
NGC\,520 &  \nodata & \nodata & 7.34 ( 1)  &155 (1.54) &528 (8.19) &163 (1.77) &263 (6.99) &93.31 \\
NGC\,660 &  \nodata & \nodata & 47.8 ( 1.78)  &428 (2.07) &1136 (15.44) &371 (3.98) &493 (4.86) &214.03 \\
NGC\,1097 &  \nodata & \nodata & 4.8 ( 0.58)  &98.5 (0.88) &282 (5) &61.5 (3.66) &129.6 (6.36) &93.61 \\
NGC\,1222 &  1.07 & \nodata & 1.89 ( 0.22)  &30.8 (0.2) &101.7 (1.33) &22.9 (0.35) &38.6 (0.76) &51.01 \\
NGC\,1365 &  0.98 & \nodata & \nodata ( \nodata)  &42.3 (0.56) &135.4 (2.54) &36.6 (1.13) &70.3 (1.5) &208.44 \\
NGC\,1614 &  2.88 & \nodata & 10.8 ( 0.78)  &63.4 (1.07) &242.4 (5.54) &44.7 (3.42) &63.7 (6.14) &145.85 \\
NGC\,2146 &  1.82 & \nodata & 55.4 ( 0.66)  &578 (1.86) &1677 (11.46) &397 (3.74) &549 (8.6) &334.08 \\
NGC\,2623 &  \nodata & \nodata & 2.41 ( 0.2)  &23.1 (0.31) &104.5 (1.35) &15.9 (0.4) &24.3 (0.39) &45.94 \\
NGC\,3310 &  1.32 & 0.33 & 4.33 ( 0.43)  &50.9 (0.36) &117.8 (1.27) &29.7 (0.25) &40.3 (0.38) &31.9 \\
NGC\,3556 &  0.6 & \nodata & 0.97 ( 0.33)  &28.3 (0.24) &81.1 (1.49) &22.1 (0.43) &27.3 (0.43) &22.28 \\
NGC\,3628 &  0.51 & 0.81 & 52.3 ( 1.26)  &375 (2.3) &994 (9.02) &279 (4.24) &333 (4.39) &89.14 \\
NGC\,4088 &  0.74 & \nodata & 2.65 ( 0.26)  &22.9 (0.39) &60.5 (1.15) &19 (0.34) &25.5 (0.88) &19.13 \\
NGC\,4676 &  1.17 & \nodata & 1.68 ( 0.11)  &10.3 (0.13) &27.53 (0.8) &\nodata  &9.06 (0.14) &11.12 \\
NGC\,4818 &  2.09 & \nodata & 8.56 ( 0.23)  &73.6 (0.41) &194.7 (1.7) &47.7 (0.61) &74 (1.23) &120.63 \\
NGC\,7714 &  1.17 & 0.4 & 3.79 ( 0.19)  &35.3 (0.29) &95.7 (1.38) &27.2 (0.45) &28.3 (0.44) &68.57 \\
\enddata
\tablecomments{Fluxes are given in $\times10^{-20}$\,W\,cm$^{-2}$. The 14\mic\ continuum flux is integrated on a 1\mic\ window centered at 14.3\mic. The uncertainty is given in parentheses. We assume an uncertainty of 10\%\ for the 14\mic\ continuum due to the instrument flux calibration. PAH fluxes were measured using PAHFIT (Sect.\,\ref{sec:obsparam_moldust}).}
\tablenotetext{a}{Metallicity estimate (in solar units) using the neon abundance derived using MIR fine-structure lines by Wu et al.\ (2008) for the BCDs and Bernard-Salas et al.\ (2009) for the starburst galaxies. The solar reference we used is $12+\log$(Ne/H)$=7.84$ (Asplund et al.\ 2006).}
\tablenotetext{b}{Metallicity estimate (in solar units) from optical studies, Bergvall \& \"Ostlin (2002) for Haro\,11, Guseva et al.\ (2000) for IIZw40, Izotov et al.\ (1994) for Mark\,1450, Shi et al.\, (2005) for Mark\,1499, Izotov \& Thuan (2004) for NGC\,1140, Kobulnicky et al.\ (1997) for NGC\,1569, Pilyugin et al.\ (2004) for IC\,342, Pastoriza et al.\ (1993) for NGC\,3310, Engelbracht et al.\ (2005) for NGC\,3628. The solar reference we used is $12+\log$(Ne/H)$=7.84$ and $12+\log$(O/H)$=8.66$ (Asplund et al.\ 2006).}
\end{deluxetable*}

In the following, we compare the PAH emission to that of the dust grains responsible for the continuum emission at wavelengths longward of 10\mic. The grain emission is determined by integrating the continuum flux in a 1\mic\ wide window centered at 14.3\mic, where there is no contamination by lines or molecular features. The continuum flux is given in Table\,\ref{tab:pahs}. Close to the ionizing sources, the continuum emission is expected to be dominated by warm big grains ($\gtrsim100$\AA\ radius) in thermal equilibrium with the ISRF. At larger distances, i.e., at PDRs, dust particles are colder and there might be a contribution of smaller grains ($10$-$100$\AA) stochastically heated (D\'esert et al.\ 1990). Note however that this paradigm is modulated by the ISRF intensity and electron density that control the heating mechanism (see Sect.\,\ref{sec:model}). In NGC\,3603, Lebouteiller et al.\ (2007) observed that, unlike PAHs, warm grains traced by the $14$\mic\ continuum are able to survive the radiation field in the ionized gas. The grain emission in this region is spatially correlated with [S\4] emission, little with [Ne\2], and is anticorrelated with [Ar\2]. Accordingly, warm grains are the main dust component in the ionized gas (see also Cesarsky et al.\ 1996; Verstraete et al.\ 1996). The same result is found by Bernard-Salas et al.\ (2011) in 30\,Dor. 

Finally, we use the H$_2$ S(1) and S(2) lines, at respectively 17.03\mic\ and 12.28\mic, to trace molecular hydrogen (Table\,\ref{tab:h2temp}). The S(0) line at 28.22\mic\ is not detected toward most positions. In H\2\ nuclei, the S(0) line contributes to about 25\% of the total of S(0-2) (Roussel et al.\ 2007).

\begin{deluxetable}{lll}
\tablewidth{0pc}
\tabletypesize{\scriptsize}
\tablecaption{Molecular hydrogen line fluxes.\label{tab:h2temp}}
\tablehead{ \colhead{} & \colhead{H$_2$ S(1)} & \colhead{H$_2$ S(2)}  }
\startdata
NGC3603\#3 & 1.62 (0.60) & 1.73 (0.60)     \\
NGC3603\#4 & $<$1.57  & 1.43    (0.60)          \\
NGC3603\#5 & 0.83 (0.60)  & 1.07  (0.60)       \\
NGC3603\#6 & $<$0.38   & 3.49    (0.60)          \\
NGC3603\#7 & $<$6.28   &  3.53   (0.60)          \\
NGC3603\#8 & $<$11.4    &  $<$7.25         \\
\tableline 
30DOR\#2  & 0.37 (0.10)      & 0.41  (0.10)     \\
30DOR\#3  & 0.12  (0.10)    & $<0.11$      \\
30DOR\#4  & $<$0.20  & 0.18  (0.10)       \\
30DOR\#5 & 0.52 (0.10)     & 0.46 (0.10)      \\
30DOR\#6 & 0.50 (0.10)     &  0.44 (0.10)     \\
30DOR\#7 & 0.31 (0.10)     &  0.22 (0.10)      \\
30DOR\#8 &  0.65 (0.10)     &  0.60  (0.10)    \\
30DOR\#10 & 0.42 (0.10)      &  0.53 (0.10)    \\
30DOR\#11 & 0.36   (0.10)   &  0.41  (0.10)    \\
30DOR\#12 & 0.61  (0.10)    &  0.59  (0.10)   \\
30DOR\#13 & $<$0.10&  0.10  (0.10)    \\
30DOR\#14 & 0.68 (0.10)     & 0.60 (0.10)      \\
30DOR\#15 & 0.80 (0.10)     &  0.78 (0.10)    \\
30DOR\#16 & $<$0.40  &  $<$0.35    \\
30DOR\#17 & 0.50  (0.10)    &  0.45  (0.10)    \\
\tableline
N66\#1   & 0.03 (0.02)      &  0.02 (0.02)      \\
N66\#2   & 0.07  (0.02)    &  0.05 (0.02)       \\
N66\#3   & 0.05  (0.02)    &  0.02  (0.02)      \\
N66\#5   & 0.09 (0.02)     &  0.08  (0.02)      \\
N66\#6   & 0.15  (0.02)    &  0.12 (0.02)   \\
N66\#7   & 0.15  (0.02)    &  0.12 (0.02)  \\
N66\#8   & 0.14 (0.02)     &  0.15 (0.02)     \\
N66\#9   & 0.29  (0.02)    &  0.29  (0.02)      \\
N66\#10   & 0.29 (0.02)     & 0.29 (0.02)      \\
N66\#11   & 0.29 (0.02)  &  0.29   (0.02)       \\
N66\#12   & $<$0.07   &  $<$0.10        \\
N66\#13   & 0.10 (0.02)  &  0.13    (0.02) 
\enddata
\tablecomments{\ Line fluxes are expressed in $\times10^{-20}$\,W\,cm$^{-2}$. The uncertainty is given in parentheses.}
\end{deluxetable}

\subsubsection{Ionized gas}\label{sec:obsparam_ig}

Several forbidden emission-lines can be used to trace the warm photoionized gas and its properties (ISRF intensity and hardness). On one hand, the H\1\ recombination line 12.37\mic\ Hu$\alpha$ depends on the physical conditions (electronic density and temperature) but not on the ISRF hardness or on the metallicity. On the other hand, the sum of the 12.81\mic\ [Ne\2] and 15.56\mic\ [Ne\3] line fluxes depends on the neon abundance but not on the radiation hardness as Ne\2\ and Ne\3\ are the dominant ionization stages of neon in the warm ionized gas. The quantity we use in the following as a proxy for the excited neon flux is [Ne\2] + $C$\,[Ne\3] where $C$ accounts for the fact that the 2 lines have different spontaneous transition rates and different level populations:
\begin{equation}
C = \frac{ \lambda_{[NeIII]} }{ \lambda_{[NeII]} } \cdot \frac{ A_{[NeII]} }{ A_{[NeIII]} } \cdot \frac{ g_{[NeII]} }{ g_{[NeIII]} },
\end{equation}
where, for a given transition, $\lambda$ is the wavelength, $A$ is the Einstein coefficient, and $g$ is the ratio of the population of the level from which the line originates to the total population of the ion. Results from Lebouteiller et al.\ (2008) for the giant H\2\ regions and Bernard-Salas et al.\ (2009) for the starburst galaxies indicate that $C\approx0.48\pm0.01$, the deviation being dominated by slightly different electron densities and temperature. The quantity [Ne\2] + $C$ [Ne\3] is directly proportional to the number of Ne\2\ and Ne\3\ ions. 

Line intensities in the giant H\2\ regions are from Lebouteiller et al.\ (2008). The [Ne\3] line could not be measured in positions \#7 and \#8 in NGC\,3603. For \#7 we estimated the [Ne\3] flux using the correlation between [Ne\3]/[Ne\2] and [S\4]/[S\3] (Beir{\~a}o et al.\ 2006; Bernard-Salas et al.\ 2009). For position \#8 we used an empirical relation between [S\4]/[Ne\2] and [Ne\3]/[Ne\2] ratios (this study and Groves et al.\ 2008). For the star-forming galaxies, we use the line measurements from Wu et al.\ (2006) for the BCDs, Bernard-Salas et al.\ (2009) for the starburst galaxies, and Smith et al.\ (2007) for H\2\ galaxies. H\1\ line measurements for BCDs were provided by Yanling Wu (private communication).

\subsubsection{Radiation field}\label{sec:obsparam_rf}

\begin{figure}
\centering
\includegraphics[angle=0,scale=0.5,clip=true]{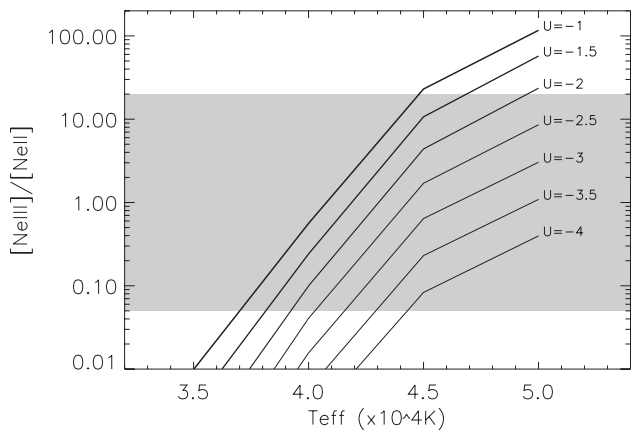}
\figcaption{[Ne\3]/[Ne\2] predicted by CLOUDY as a function of the radiation hardness and of the ionization parameter (tracks). See text for model details. The gray zone indicates the range of line ratios observed in the objects of this study. \label{fig:cloudy}}
\end{figure}

The gas excitation is usually probed by line ratios of adjacent ionization states, such as [Ne\3]/[Ne\2] or [S\4]/[S\3]. These ratios are proportional to the radiation hardness with also a $-$ smaller $-$ dependence on the ionization parameter $U$ (e.g., Morisset et al.\ 2004). We ran a grid through the photoionization code CLOUDY (version c08.00; Ferland et al.\ 1998) to estimate the [Ne\3]/[Ne\2] ratio in various conditions. A single star was chosen with a temperature $T$ between 25\,000\,K and 50\,000\,K, the ionization parameter $U$ spanning from $-4$ to $-1$, and the inner radius to the illuminated slab spanning from 0.1 to 1\,parsec. Results are plotted in Fig.\,\ref{fig:cloudy}. We find that the [Ne\3]/[Ne\2] ratio is proportional to $T \times U^{0.024}$ and is thus mostly dependent on the star temperature hence the ISRF hardness.

We favor hereafter the line ratio [Ne\3] 15.56\mic/[Ne\2] 12.81\mic\ which is little affected by silicate dust extinction. Moreover, the two lines arise in the same module of the \textit{IRS} so that flux scaling is not needed. Since the factor $C$ is essentially the same for all objects (Sect.\,\ref{sec:obsparam_ig}), we ignore it for the neon line ratio.

\subsubsection{Chemical abundances}\label{sec:obsparam_met}

The metallicity is derived from the neon abundance calculated using MIR fine-structure lines (Wu et al.\ 2008; Bernard-Salas et al.\ 2009). This choice was motivated by the fact that about half of the galaxies we consider do not have referenced abundance measurements from optical spectra. Moreover, IR abundance determinations are less affected by dust extinction. Our results remain unchanged by using the optically derived abundances (Sect.\,\ref{sec:metals_sfg}). Metallicity values are listed in Table\,\ref{tab:sfg}. 

\section{Giant H\2\ regions}\label{sec:destr_ghiir}

\subsection{Physical conditions}\label{sec:physcond}

\begin{figure}
\centering
\includegraphics[angle=0,scale=0.5,clip=true]{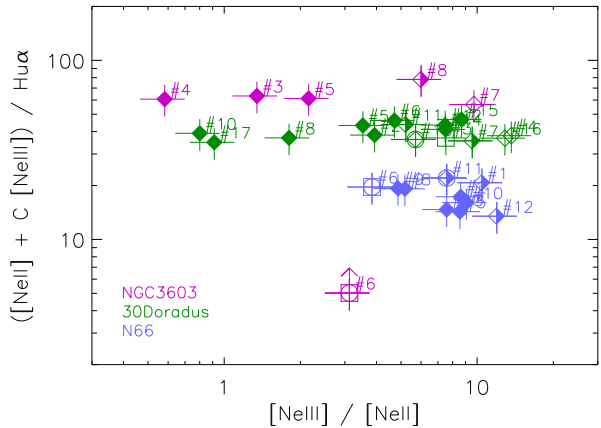}
\figcaption{The quantity ([Ne\2]+$C$[Ne\3])/Hu$\alpha$ (which is proportional to the neon abundance, see text) is plotted against the gas excitation as traced by the neon line ratio [Ne\3]/[Ne\2]. Squares indicate ionizing clusters, circles regions with silicate in emission, filled diamonds PDRs, open-diamonds ionized gas, and half-filled diamonds regions showing both PDR and ionized gas tracers (see Sect.\,\ref{sec:obsparam} and Lebouteiller et al.\ 2008 for the spectra and the main spectral characteristics).\label{fig:isrf2}}
\end{figure}

The observations of the three giant H\2\ regions span a wide range of metallicity, while at the same time probing different physical conditions within each region (Fig.\,\ref{fig:isrf2}). N\,66 is characterized by a globally harder ISRF than the two other regions. Furthermore, although NGC\,3603 and 30\,Dor are significantly more metal-rich than N\,66, they do contain high-excitation gas.

\subsection{PAH vs.\ warm dust}\label{sec:rfp}

\begin{figure}
\includegraphics[angle=0,scale=0.5,clip=true]{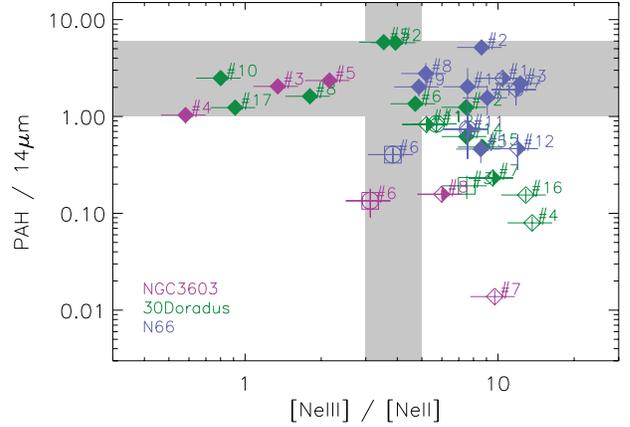}
\figcaption{The PAH emission measured against the 14\mic\ dust continuum is plotted as a function of the ISRF hardness as defined in Sect.\,\ref{sec:obsparam_rf}.  The horizontal gray band illustrates the location of positions with significant PAH emission, while the vertical band indicates the threshold above which PAHs are believed to be destroyed by energetic photons. See Fig.\,\ref{fig:isrf2} for the symbol explanation. 
\label{fig:pah_isrf_ghiir}}
\end{figure}

We plot in Fig.\,\ref{fig:pah_isrf_ghiir} the PAH/14\mic\ ratio as a function of the ISRF hardness in the three giant H\2\ regions of our sample. As explained in detail in the following, we distinguish 2 behaviors depending on the ISRF hardness. For a soft radiation field, the PAH/14\mic\ ratio is fairly independent on the IRSF hardness and shows a dispersion of less than a factor 5. For a hard radiation field, PAH/14\mic\ shows a much wider dispersion with values that are systematically equal to or lower than the average ratio observed in a soft radiation field environment. 

Although not shown, we obtain a fairly similar trend when analyzing the sum of the PAH equivalent widths rather than the PAH/14\mic\ flux ratio. This is because the dust continuum lying under the PAH features is on first order correlated with the continuum at 14\mic. By normalizing the strength of the PAH features with the 14\mic\ continuum, we aim at minimizing systematic effects due to the fact that the dust continuum between 6\mic\ and 14\mic\ may show variations (different dust temperature, presence of silicate dust emission/absorption, stochastic heating of dust grains vs.\ thermal equilibrium).

\subsubsection{Stellar clusters}

The central stellar clusters (position \#6 in NGC\,3603, \#3 in 30\,Dor, and \#6 in N\,66) lie somewhat off the trend identified in Sect.\,\ref{sec:rfp}. They display low PAH/14\mic\ ratios $-$ which is expected since PAHs are likely destroyed by photons from massive stars in the cluster $-$ but the ISRF hardness is not the highest in the sample. This was also observed in NGC\,3603 at small spatial-scales by Lebouteiller et al.\ (2007) who concluded that the presence of a cavity around the central stellar cluster (volume in which [Ne\3] and [S\4] would strongly emit) cleared out by massive stars can explain this effect. In the case of N\,66, the most massive stars are actually not in the cluster NGC\,346 and the ISRF hardness peaks to the east of it (Whelan et al.\ 2010).

\subsubsection{Soft radiation field environment}

The spectra toward all the positions characterized by soft radiation ([Ne\3]/[Ne\2]$\lesssim3$) are PAH-dominated and correspond to individual PDRs (filled diamonds in Fig.~\ref{fig:pah_isrf_ghiir}). The dispersion of the PAH/14\mic\ ratio along the "PDR stripe" (around a factor 5, indicated by the horizontal band) is much smaller than the global variations between all the positions ($\approx3$\,dex), implying that the PAHs and the dust particles responsible for the continuum emission scale with each other on first order in PDRs. For a soft radiation field, one expects the dust continuum in PDRs to be dominated by stochastically heated grains (very small grains; VSGs) (see e.g., Rapacioli et al.\ 2006; Bern{\'e} et al.\ 2007). Given the size of the extraction window of the closest giant H\2\ region NGC\,3603 (Sect.\,\ref{sec:obs_ghiir}), this suggests that PAHs and VSGs emit in a zone of  less than $0.4$\,pc across (projected). 

Sources with significant 14\mic\ emission unrelated to the present star-formation episode are expected to lie well below the PDR stripe (horizontal band in Fig.\,\ref{fig:pah_isrf_ghiir}), while sources with significant PAH contamination from excitation by older stellar populations are expected to lie above. The upper limit on the potential contamination is given by the dispersion along the PDR stripe, i.e., a factor 5. 
Since the molecule and dust excitation in giant H\2\ regions is dominated by massive stars, we consider the PDR stripe as a reference for interpreting integrated emission of galaxies where other sources of excitation can contribute (Sect.\,\ref{sec:sfg}). 

\subsubsection{Hard radiation field environment}\label{sec:hardrfe}

Sources with low PAH/14\mic\ ratios are systematically characterized by hard radiation in the ionized gas ([Ne\3]/[Ne\2]$\gtrsim3$; Fig.\,\ref{fig:pah_isrf_ghiir}). The PAH and the dust emission thus become decoupled in extreme environments. Most of the positions with low PAH/14\mic\ ratios were identified as pure ionized gas close to the central clusters by Lebouteiller et al.\ (2008) using both MIR imagery and spectroscopy (empty and half-filled diamonds in Fig.~\ref{fig:pah_isrf_ghiir}). In the ionized gas, the dust continuum at 14\mic\ is expected to be dominated by big grains in thermal equilibrium with the radiation field. 

For a given giant H\2\ region, observed variations in the PAH/14\mic\ ratio are not controlled by metallicity effects since metal abundances are found to be remarkably uniform (Lebouteiller et al.\ 2008). 
The drop seen around [Ne\3]/[Ne\2]$\gtrsim3$ is due to a combination of two effects: the transition from VSGs in the PDRs to big grains in the ionized gas which results in increasing the 14\mic\ continuum level (see also Dale et al.\ 2000), and the transition from PAH excitation in the PDRs to PAH destruction in the ionized gas. 
Although both effects are degenerate in Figure~\ref{fig:pah_isrf_ghiir}, a variation of the intrinsic PAH abundance is necessary to explain the observed variations traced by the PAH/14\mic\ ratio (see Sect.\,\ref{sec:model} and Fig.\,\ref{fig:model2}). Destruction of PAHs is most likely responsible for the low PAH abundance in the ionized gas, as it has already been observed in Galactic H\2\ regions  (e.g., Geballe et al.\ 1989; Verstraete et al.\ 1996; Cesarsky et al.\ 1996; Madden et al.\ 2006). Although it is also possible that the density profile controls the PAH abundance if the latter results from an equilibrium between thermal evaporation and reconstruction by encounters with C$^+$ (Giard et al.\ 1994), our results suggest that the ISRF hardness plays an important role in the molecule destruction (see also the comparison between PAH and H$_2$ emission in Sect.\,\ref{sec:molh}).

\subsection{Dust emission properties in N\,66}\label{sec:destr_ghiir_met}

\begin{figure}
\includegraphics[angle=0,scale=0.5,clip=true]{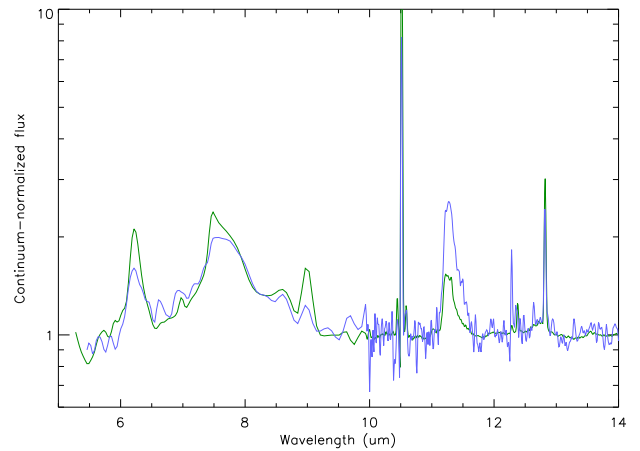}
\figcaption{The positions \#14 and \#15 in 30\,Dor (combined - green) show the closest continuum-normalized MIR spectrum to that of the position \#10 in N\,66 (blue). The continuum was removed using a spline function. The 11.3\mic\ PAH band across N\,66 is consistently stronger relative to the other bands as compared to the other H\2\ regions. \label{fig:compan6630dor}}
\end{figure}

The PAH/14\mic\ values in N\,66 are larger than what could be expected from the radiation field properties, since all the sources in this object have [Ne\3]/[Ne\2]$\gtrsim4$ (Fig.~\ref{fig:pah_isrf_ghiir}). Figure\,\ref{fig:compan6630dor} shows that continuum-normalized spectra of typical PDR-dominated sources in N\,66 (\#10) and 30\,Dor (\#14, \#15) have similar PAH integrated strength. It must be reminded here that the aperture physical size is almost identical for the observations in N\,66 and 30\,Dor as these objects lie almost at the same distance (Sect.\,\ref{sec:obs_ghiir}). Although the PAH band ratios differ in N\,66 and 30\,Dor (see Galliano et al.\ 2008a), the total PAH luminosity is identical. Hence the fact that PAH/14\mic\ values in N\,66 are offset with respect to 30\,Dor sources is due to relatively reduced warm dust emission in N\,66 rather than enhanced PAH emission. Could the weak dust emission in N\,66 be a direct effect of metallicity on the dust grain abundance? The dust emission in the ionized gas is indeed expected to be reduced in low-metallicity environments, since the dust-to-gas ratio scales directly with the ISM metallicity (Leroy et al.\ 2007; Lisenfeld \& Ferrara 1998). If PAH emission would follow the same metallicity-dependence as dust emission, one would expect the PAH/14\mic\ ratio in N\,66 and 30\,Dor to be identical for a given ISRF hardness. Our results therefore seem to imply that PAH emission and dust emission do not scale together with metallicity and, more to the point, that PAH emission would depend less on metallicity than the warm dust emission. This is however difficult to reconcile with models of dust formation (Sect.\,\ref{sec:form}).

The relatively low metallicity of N\,66 could have an indirect influence via a change in the physical conditions leading to enhanced molecule photodestruction. Metal-poor sources are usually characterized by a relatively harder ISRF (Sect.\,\ref{sec:destr}) and also have a more intense ISRF because of the relatively low dust-to-gas ratio which reduces the ISM opacity. Nevertheless, NGC\,3603 and 30\,Dor, which are more metal-rich than N\,66, also contain regions with a comparable excitation degree (Sect.\,\ref{sec:obsparam_ig}). Therefore, if metallicity had only an indirect influence through variations of the ISRF properties, data points of the H\2\ regions should align with each other in Fig.\,\ref{fig:pah_isrf_ghiir}. 

Besides metallicity, a shift in the PAH/14\mic\ ratio can be observed because of different ISRF properties affecting the dust heating mechanism. For a given ISRF hardness ([Ne\3]/[Ne\2]), a lower dust temperature implies a higher PAH/14\mic\ (Sect.\,\ref{sec:model}).

\subsection{Theoretical influence of the ISRF properties on the PAH/14\mic\ ratio}\label{sec:model}

Independently from metallicity and the possible contribution of different phases within a given beam, the simple variations of the illumination conditions of the PAH and dust grains can cause significant changes in the observed PAH/14\mic\ ratio. In order to quantify these variations, we have modeled realistic dust spectra with a homogeneous ISRF. We assumed the dust size distributions and abundances of the diffuse Galactic ISM modeled by Zubko et al.\ (2004), assuming bare grains with solar abundance constraints. This grain mixture is made of PAH (neutral and cationic; Draine \& Li 2007), graphite (Laor \& Draine 1993), and silicate (Weingartner \& Draine 2001). The temperature fluctuations are computed using the method of Guhathakurta \& Draine (1989). 

In order to parametrize the ISRF, $J_\lambda$, we synthesize a single instantaneous burst with the stellar population synthesis model PEGASE (Fioc \& Rocca-Volmerange 1997). We measure the synthesized ISRF at 0\,Myr. Since PAHs emit principally from neutral regions (PDRs and diffuse ISM), we cut out the ionizing continuum of this ISRF. The family of ISRFs is parametrized by two quantities, its energy density:
\begin{equation}
  \mathcal{U} = \int_{0.0912\;\mu m}^{8\;\mu m}J_\lambda d\lambda,
\end{equation}
and its hardness:
\begin{equation}
  \langle h\nu\rangle = \frac{\displaystyle \int_{0.0912\;\mu m}^{8\;\mu m}J_\lambda d\lambda}{\displaystyle \int_{0.0912\;\mu m}^{8\;\mu m}J_\lambda/(hc/\lambda) d\lambda}.
\end{equation}

The energy density is a simple scaling factor of the ISRF. The hardness parameter quantifies the weight of the UV photons compared to the redder ones. There are many ways to modify the hardness of an ISRF. We chose to apply a screen extinction $\exp(-\tau(\lambda))$, varying the optical depth. This choice is purely formal. High optical depths make the ISRF redder and decrease the subsequent $\langle h \nu\rangle$ parameter. 

\begin{figure}
\includegraphics[angle=0,scale=0.6,clip=true]{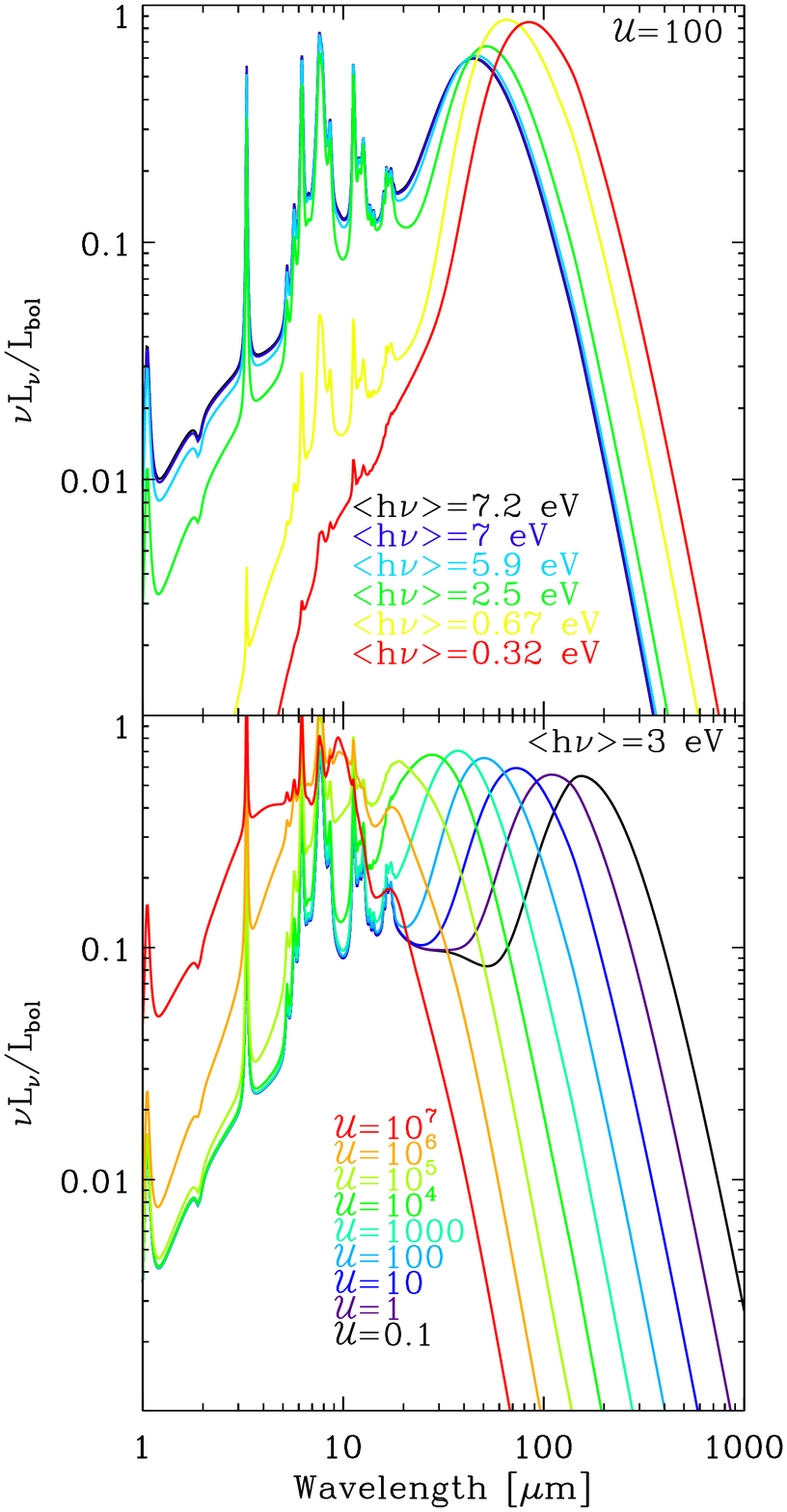}
\figcaption{Dust spectra used to study the effects of the illumination conditions on the PAH/14\mic\ ratio, from a purely theoretical point of view. Both panel represent families of dust spectra exposed to a single ISRF. Each spectrum is normalized by its bolometric luminosity. The incident ISRF is parametrized by its total energy density $\mathcal{U}$ and by its hardness. The top panel demonstrates the effect of the hardness, fixing $\mathcal{U}$. The bottom panel demonstrates the effect of the starlight intensity, fixing the hardness.\label{fig:model1}}
\end{figure}

The dust spectra generated are shown in Fig.\,\ref{fig:model1}. The top panel of Fig.\,\ref{fig:model1} demonstrates the effect of the ISRF hardness on the shape of the dust SED. Hard ISRFs have prominent PAH features. When the hardness decreases the PAH strength decreases, as the UV-to-visible ratio decreases. This is due to the fact that PAHs have a much steeper cross-section in the UV, compared to other grain species (e.g., Fig.\,2 of Galliano et al. 2008b). 

\begin{figure}
\includegraphics[angle=0,scale=0.42,clip=true]{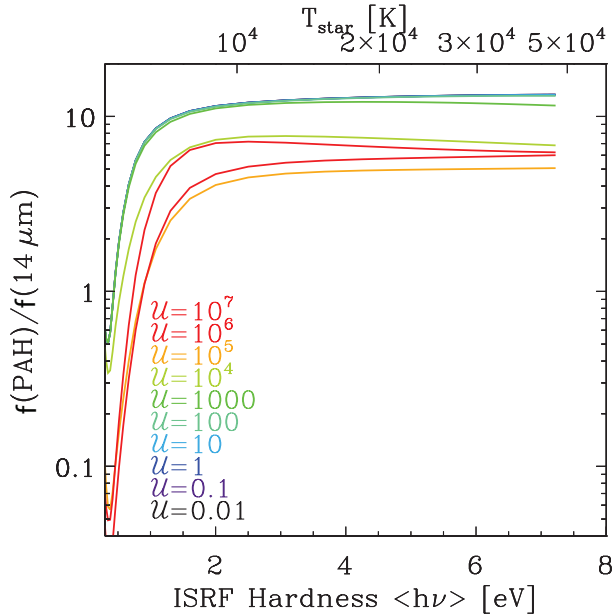}
\figcaption{Theoretical variations of the PAH/14\mic\ ratio, as a function of the illumination conditions. Those curves are a directly integrated from the SEDs of Fig.\,\ref{fig:model1}. $\mathcal{U}$ is the energy density of the ISRF, and $\langle h\nu\rangle$ its hardness. The top axis gives the effective temperature of a single O or B star that would have the hardness of the bottom axis.\label{fig:model2}}
\end{figure}

On the other hand, the bottom panel of Fig.\,\ref{fig:model1} shows the variations of the shape of the dust spectrum with the energy density. In this case the hardness of the ISRF is fixed. PAHs are stochastically heated. They are controlled by single photon heating. It means that the shape of their emission spectrum will depend only on the average energy of the incident photons ($\langle h\nu\rangle$), and not on their rate ($\mathcal{U}$). The very small grains, responsible for the flat continuum between 20\,\mic\ and 40\,\mic\ , for $\mathcal{U}\lesssim10$ on the bottom panel of Fig.\,\ref{fig:model1}, are in the same regime. However, when the energy density increases, the equilibrium temperature of the large grains increases, and the largest small grains reach thermal equilibrium. Consequently the emission peak shifts at shorter wavelengths. For high energy densities ($\mathcal{U}\gtrsim 10^3$), the grains responsible for the 14\,\mic\ emission are mainly at thermal equilibrium. Therefore, their emissivity\footnote{We refer to the monochromatic emissivity $\epsilon_\nu (\nu)$ (per unit frequency, per unit mass, averaged over solid angle) of dust grains at thermal equilibrium with the ISRF. In this context, $\epsilon_\nu (\nu) = \kappa (\nu) \times B_\nu(T_{eq}, \nu)$, where $\kappa (\nu)$ is the grain opacity and the $B_\nu(T_{eq}, \nu)$ is the Planck function for an equilibrium temperature $T_{eq}$. } is enhanced, and the PAH/14\mic\ ratio decreases.

Figure\,\ref{fig:model2} shows the effect of varying $\langle h\nu\rangle$ and $\mathcal{U}$ on the PAH/14\mic\ ratio. For comparison with our measurements, the flux of the PAHs, $f$(PAH), includes all the bands between 5\,\mic\ and 12\,\mic. The 14\,\mic\ flux, $f$(14\,\mic), is the dust continuum integrated between 13.8\,\mic\ and 14.8\,\mic. We have indicated, on the top axis, the hardness corresponding to a single star (black body), with an effective temperature $T_{\rm star}$. Figure\,\ref{fig:model2} demonstrates that, in the range corresponding to the regions studied in this paper ($T_{\rm star}>10^4$\,K, or $\langle h\nu\rangle \gtrsim 2$; Fig.\,\ref{fig:cloudy}), the spread in the PAH/14\mic\ ratio due to variations in the illumination conditions is at most a factor of $\approx2.6$. It is therefore much lower than the observed spread ($\sim100$), which consequently requires variations of the intrinsic abundance of PAHs. Our results suggest that PAH destruction in the ionized gas could explain the observed variations of the PAH/14\mic\ ratio across each region (Sect.\,\ref{sec:hardrfe}).

\subsection{PAHs vs.\ ionized gas}\label{sec:transition}

We next compare PAH emission to the H\1\ recombination line Hu$\alpha$ which traces the warm photoionized gas  (Sect.\,\ref{sec:obsparam_ig}). While the 14\mic\ continuum can be dominated by a different dust component in PDRs and in the ionized gas, Hu$\alpha$ simply provides a tracer of the ionized gas. As a result, the PAH/Hu$\alpha$ ratio is thus proportional to the relative contribution of PDR and ionized gas within the extracted beam. 

Although the quantity PAH/Hu$\alpha$ is a priori expected to depend on metallicity to first order because PAH abundance itself scales with carbon abundance, our results show that the 14\mic\ dust continuum seems to be depend more on metallicity than the PAH emission (Sect.\,\ref{sec:rfp}). We therefore choose not to normalize the PAH/Hu$\alpha$ ratio by the metallicity. Further discussion on the influence of metallicity on the dust continuum emission can be found in Sect.\,\ref{sec:metals_sfg}.

\begin{figure}
\includegraphics[angle=0,scale=0.5,clip=true]{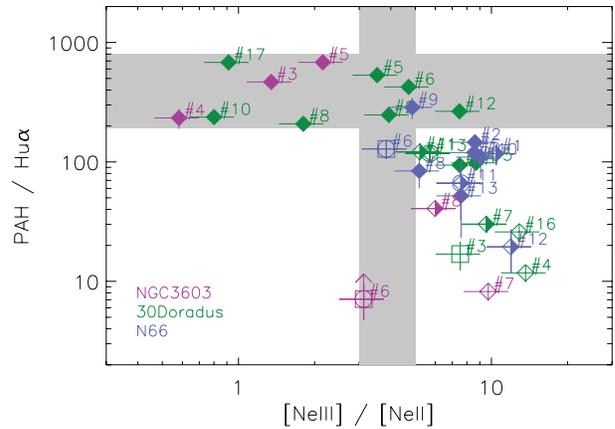}
\figcaption{The PAH emission (measured against Hu$\alpha$) is plotted as a function of the ISRF hardness. See Fig.\,\ref{fig:isrf2} for symbol explanation. 
\label{fig:pahhi_isrf_ghiir}}
\end{figure}

The PAH/Hu$\alpha$ ratio shows a smooth trend as a function of the ISRF hardness (Fig.\,\ref{fig:pahhi_isrf_ghiir}). 
The main difference between the PAH/Hu$\alpha$ ratio and the PAH/14\mic\ ratio (Fig.\,\ref{fig:pah_isrf_ghiir}) concerns the uniformity of the trend for the three giant H\2\ regions as a function of the ISRF hardness using PAH/Hu$\alpha$. In particular, the sources in N\,66 now cluster with those in NGC\,3603 and 30\,Dor, in agreement with Sect.\,\ref{sec:destr_ghiir_met} in that for the same physical conditions the 14\mic\ dust continuum in this object is fainter relatively to the other giant H\2\ regions. This result implies that the three giant H\2\ regions follow the same transition from PDR dominated sources (filled diamonds in Fig.\,\ref{fig:pahhi_isrf_ghiir}) to ionized gas dominated sources (empty diamonds) regardless of the environment metallicity. 

Since in star-forming regions PAHs and Hu$\alpha$ trace respectively the PDRs and the ionized gas, we expect the PAH/Hu$\alpha$ to reflect the mixture PDR/ionized gas. For a soft radiation field, the PAH/Hu$\alpha$ ratio seems to reach a constant value. We attribute this effect to the size of the extraction window which makes it impossible to probe "pure" exposed PDRs with no ionized gas counterparts (such as in NGC\,7023; Cesarsky et al.\ 1996), even in the closest object of our sample, NGC\,3603 (Sect.\,\ref{sec:obs_ghiir}). The examination of individual spectra confirms that none of the regions resembles an exposed PDR (Lebouteiller et al.\ 2008). Although it is clear that ionized gas always mixes along the lines of sight from the observations of lines such as Hu$\alpha$, [Ne\2], or even [Ne\3], it is difficult to assert what fraction of it is physically associated with the PDRs probed. Nevertheless, we notice that PAH dominated regions in 30\,Dor do not show systematically larger PAH/Hu$\alpha$ ratios than in NGC\,3603 despite the larger physical aperture, indicating that contamination by ionized gas in the beam is not significant (or that the contaminations by surrounding regions to the PAH and Hu$\alpha$ emission are identical). 

\subsection{Influence of shocks}\label{sec:shocks}

Shock conditions might modify the PAH abundance locally (Micelotta et al.\ 2010). O'Halloran et al.\ (2006) suggested that PAHs in low-metallicity dwarf galaxies could be destroyed by SN blast waves, based on the use of the [Fe\2]/[Ne\2] ratio. In shocked regions, the iron locked onto dust grains is released into the ISM, increasing the iron abundance. 
As noticed by O'Halloran et al.\ (2006), large [Fe\2]/[Ne\2] ratios in low-metallicity environments can also be due to the fact that the amount of iron trapped in dust grains depends on the dust-to-gas ratio which itself depends on metallicity (e.g.; Lisenfeld \& Ferrara 1998). 

\begin{figure}
\includegraphics[angle=0,scale=0.5,clip=true]{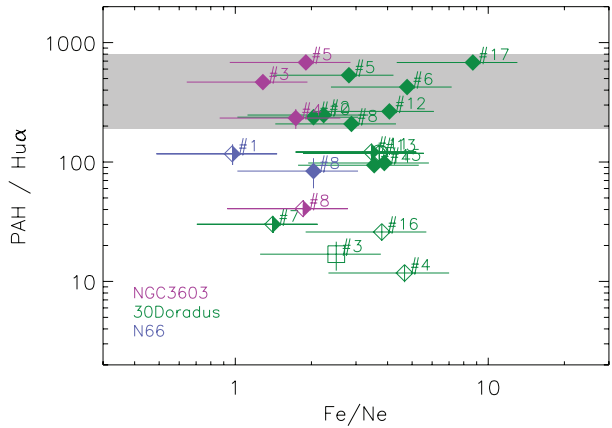}
\figcaption{The PAH/Hu$\alpha$ ratio is plotted as a function of the abundance ratio Fe/Ne. See Fig.\,\ref{fig:isrf2} for symbol explanation.\label{fig:pah_isrf_ghiir_feab}}
\end{figure}

In order to explore whether shocks can modify the PAH emission in giant H\2\ regions, we use the Fe/Ne abundance ratio instead of [Fe\2]/[Ne\2] to avoid possible biases from ionization effects, although the iron abundance determinations also bear significant errors driven by the ionization correction factors. Abundances are taken from Lebouteiller et al.\ (2008). In Figure~\ref{fig:pah_isrf_ghiir_feab}, we plot PAH/Hu$\alpha$ against the Fe/Ne ratio. It can be seen that there is no clear trend of decreasing PAH/14\mic\ with increasing Fe/Ne.

The Fe/Ne ratio toward position \#17 (corresponding to a young stellar object under the influence of nearby massive stars and a supernova) is much larger than the values toward the other sources in this region. Nevertheless, the corresponding PAH/14\mic\ ratio toward this object is among the highest in our sample, indicating that even though shocks are present, they do not result in significant PAH destruction. This result agrees with Indebetouw et al.\ (2009) who found that the ionization structure of 30\,Dor is dominated by photoionization rather than shocks. The wide range of Fe/Ne in this object is likely driven by uncertainties on the abundance determination rather than by different shock conditions (Lebouteiller et al.\ 2008).

We would like to stress that our results suffer from the fact that the PAH emission might not be associated with the shock itself (if present). For this reason, it seems also difficult to use MIR line ratios which are known to provide good constraints on shock conditions, such as [O\4]25.89\mic/[Ne\2]12.81\mic, or [Fe\2]25.98\mic/[O\4]25.98\mic\ (Allen et al.\ 2008).

\section{Star-forming galaxies}\label{sec:sfg}

The PAH equivalent width in star-forming galaxies has been shown to decrease in the presence of hard ISRF, using \textit{ISO} (Madden 2000; Madden et al.\ 2006; Thuan et al.\ 1999) and \textit{Spitzer} (Houck et al.\ 2004; Engelbracht et al.\ 2005; Wu et al.\ 2006). In this section, we examine consistent measurements of a sample of star-forming galaxies to compare to the results we obtained for the giant H\2\ regions.

\subsection{PAH vs.\ warm dust}\label{sec:isrf_sfg14}

\begin{figure}
\includegraphics[angle=0,scale=0.5,clip=true]{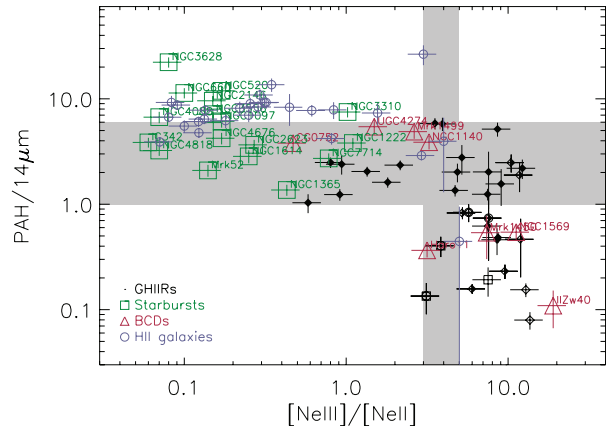}
\figcaption{The PAH/14\mic\ flux ratios is plotted as a function of the ISRF hardness as traced by [Ne\3]/[Ne\2]. The plot for the giant H\2\ regions alone is presented in Figure\,\ref{fig:pah_isrf_ghiir}. 
\label{fig:pah_isrf}}
\end{figure}

We plot in Figure~\ref{fig:pah_isrf} the relation between PAH/14\mic\ and the ISRF hardness for starburst galaxies, H\2\ galaxies, and BCDs together with those of the giant H\2\ regions. Galaxy samples are presented in Sect.\,\ref{sec:obs_ghiir}. Starburst galaxies and H\2\ galaxies are characterized by relatively soft radiation, [Ne\3]/[Ne\2]$<3$, which is expected given their high metallicity (Sect.\,\ref{sec:destr}). In contrast, BCDs, as they are more metal-poor than starburst galaxies, have relatively harder radiation ($1.5<$[Ne\3]/[Ne\2]$<18$). 

Overall, the PAH/14\mic\ in starburst galaxies and H\2\ galaxies is equal or higher than values in PDRs of giant H\2\ regions, being consistent with most of their MIR brightness originating from PDRs. Most of the dispersion is likely driven by contributions from other galaxy components (ISM, embedded compact H\2\ regions, exposed PDRs, and reflection nebula\ae) to the PAH and dust emission (see also Peeters et al.\ 2004). 

The PAH/14\mic\ in BCDs spans all the range from PDR values to ionized gas values (Sect.\,\ref{sec:destr_ghiir}). 
Four BCDs have significantly low PAH/14\mic\ ratios (Mark\,1450, NGC\,1569, IIZw\,40, and Haro\,11), and they are all characterized by [Ne\3]/[Ne\2]$>3$. In the case of Haro\,11, the remarkably low PAH/14\mic\ ratio is likely related to the enhanced dust emission observed around 24\mic\ (Engelbracht et al.\ 2008; Galametz et al.\ 2009). 

\subsection{Mixture of PDR - ionized gas}\label{sec:isrf_sfgig}

\begin{figure}
\includegraphics[angle=0,scale=0.5,clip=true]{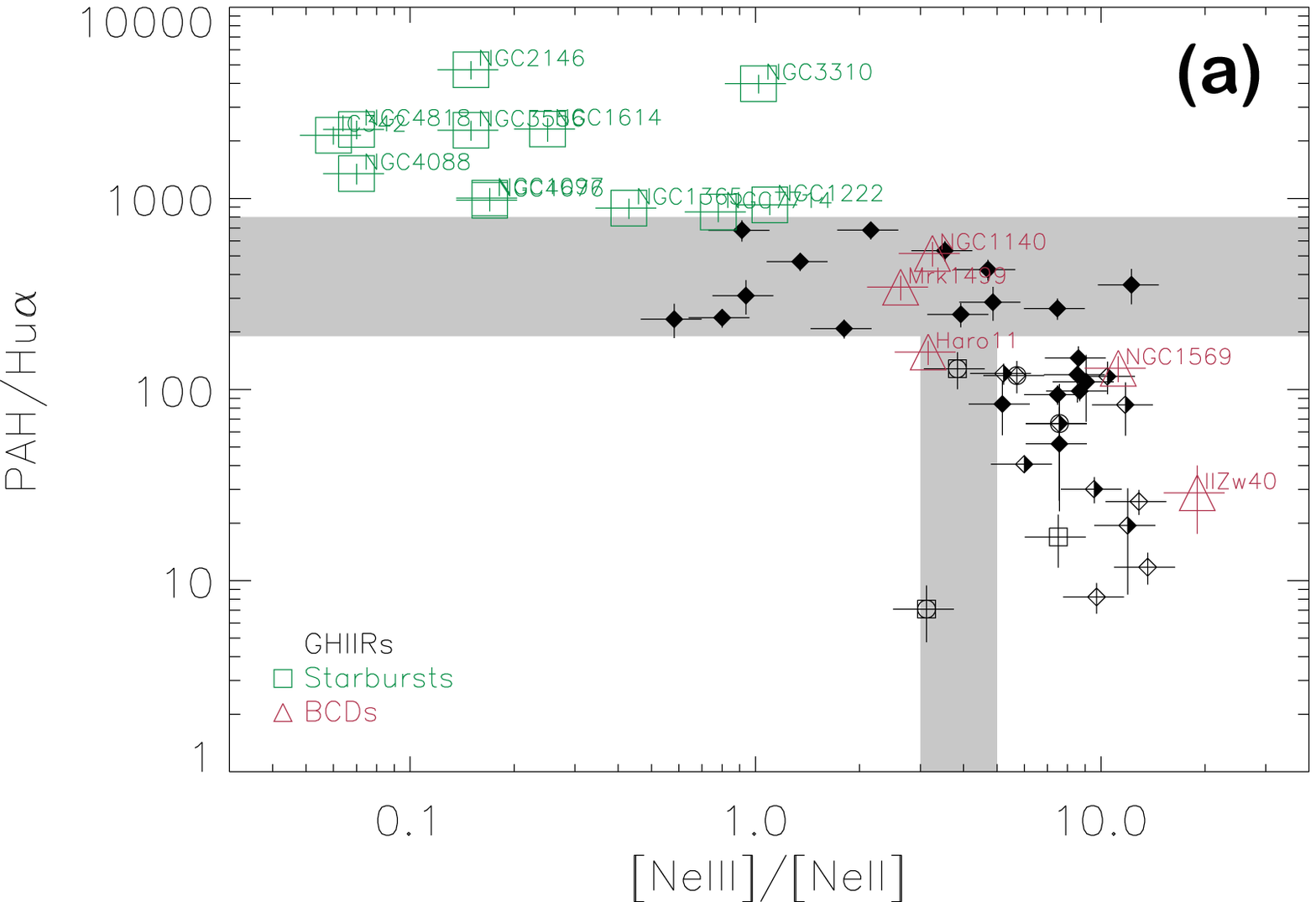}\\
\includegraphics[angle=0,scale=0.5,clip=true]{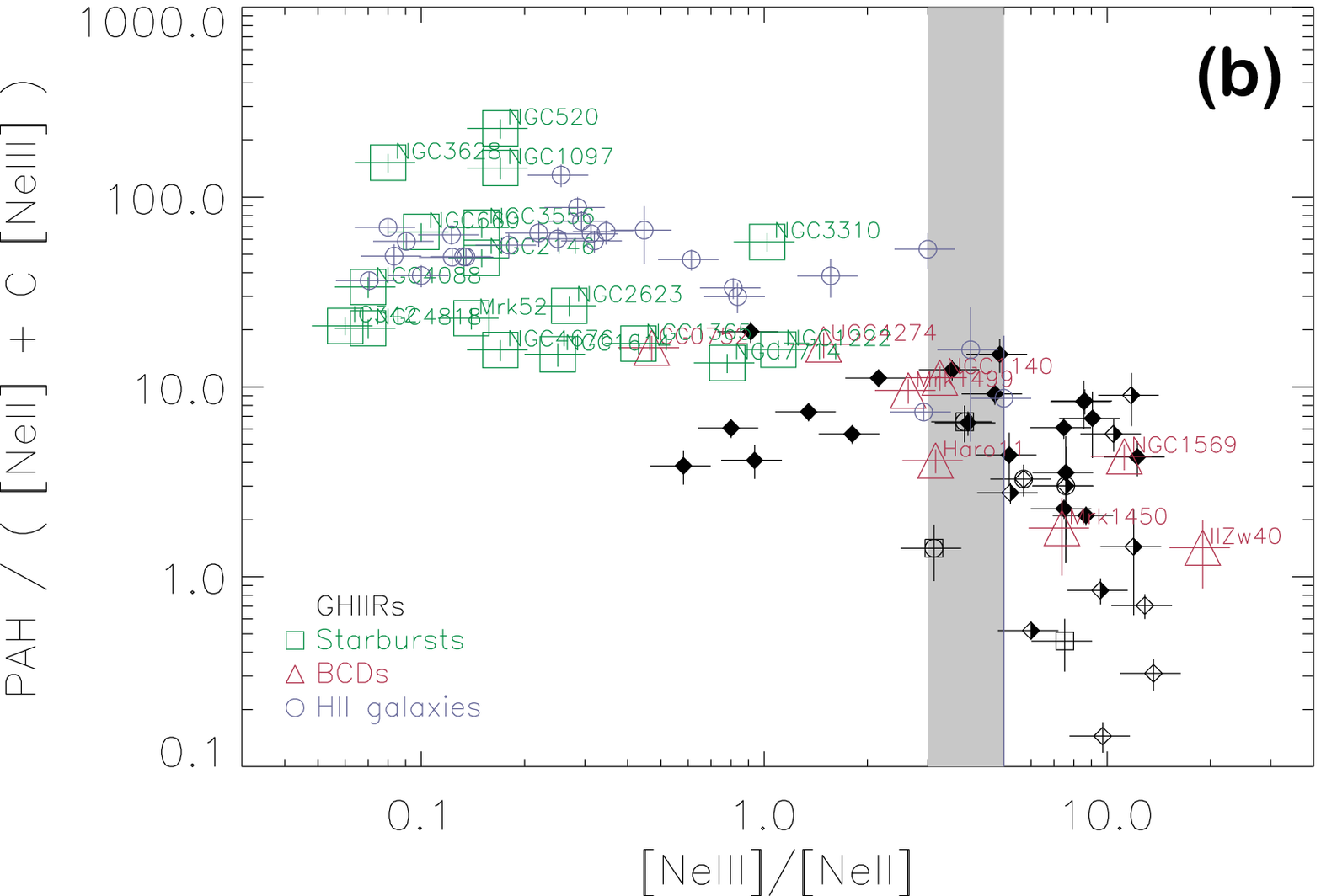}
\figcaption{The PAH/Hu$\alpha$ \textbf{(a)} and PAH/([Ne\2]+$C$[Ne\3]) \textbf{(b)} flux ratios are plotted as a function of the ISRF hardness as traced by [Ne\3]/[Ne\2]. \label{fig:pah_isrf_ig}}
\end{figure}

The PAH/Hu$\alpha$ correlation with the ISRF hardness in star-forming galaxies provides a slightly different picture (Fig.~\ref{fig:pah_isrf_ig}a). The outlier Haro\,11 in Figure~\ref{fig:pah_isrf} now lines up with the other objects like it was observed for N\,66 confirming peculiar warm dust emission in both objects (Sect.\,\ref{sec:transition}). 

Starburst galaxies seem to have systematically larger PAH/Hu$\alpha$ values than PDR sources in giant H\2\ regions. We confirm this result by using [Ne\2]+$C$[Ne\3] as a tracer of the ionized gas (Fig.\,\ref{fig:pah_isrf_ig}b). The PAH/([Ne\2]+$C$[Ne\3]) ratio should be less dependent on metallicity than PAH/Hu$\alpha$ (see also Sect.\,\ref{sec:transition}), and it allows us to increase the statistics by adding the H\2\ galaxies for which no Hu$\alpha$ measurement is available. 

These results suggest that a fraction of the PAH emission in starburst galaxies does not originate from a collection of PDRs similar to the ones observed in giant H\2\ regions. The contribution from other galaxy components probably affects significantly the PAH emission because of the large physical aperture size.
Peeters et al.\ (2004) found that the ratio between PAH luminosity and other massive star-formation tracers (Lyman continuum photons, infrared luminosity) could vary by as much as a factor 10 between H\2\ regions and integrated galaxies, which is on the same order of magnitude as the the shift in PAH/Hu$\alpha$ we observe between starbursts and individual PDRs in giant H\2\ regions. 

Most BCDs show PAH/Hu$\alpha$ ratios comparable to the values found in the ionized gas of the H\2\ regions. The morphological resemblance between MIR spectra of ionized gas in H\2\ regions and spectra of BCDs has been pointed out by several studies (e.g., Galliano et al.\ 2003; 2005; Madden et al.\ 2006; Peeters et al.\ 2004). Our results are consistent with the conditions for PAH emission and destruction in galaxies being on first order similar to those seen in giant H\2\ regions.

\subsection{Warm dust emission properties}\label{sec:metals_sfg}

\begin{figure}
\includegraphics[angle=0,scale=0.5,clip=true]{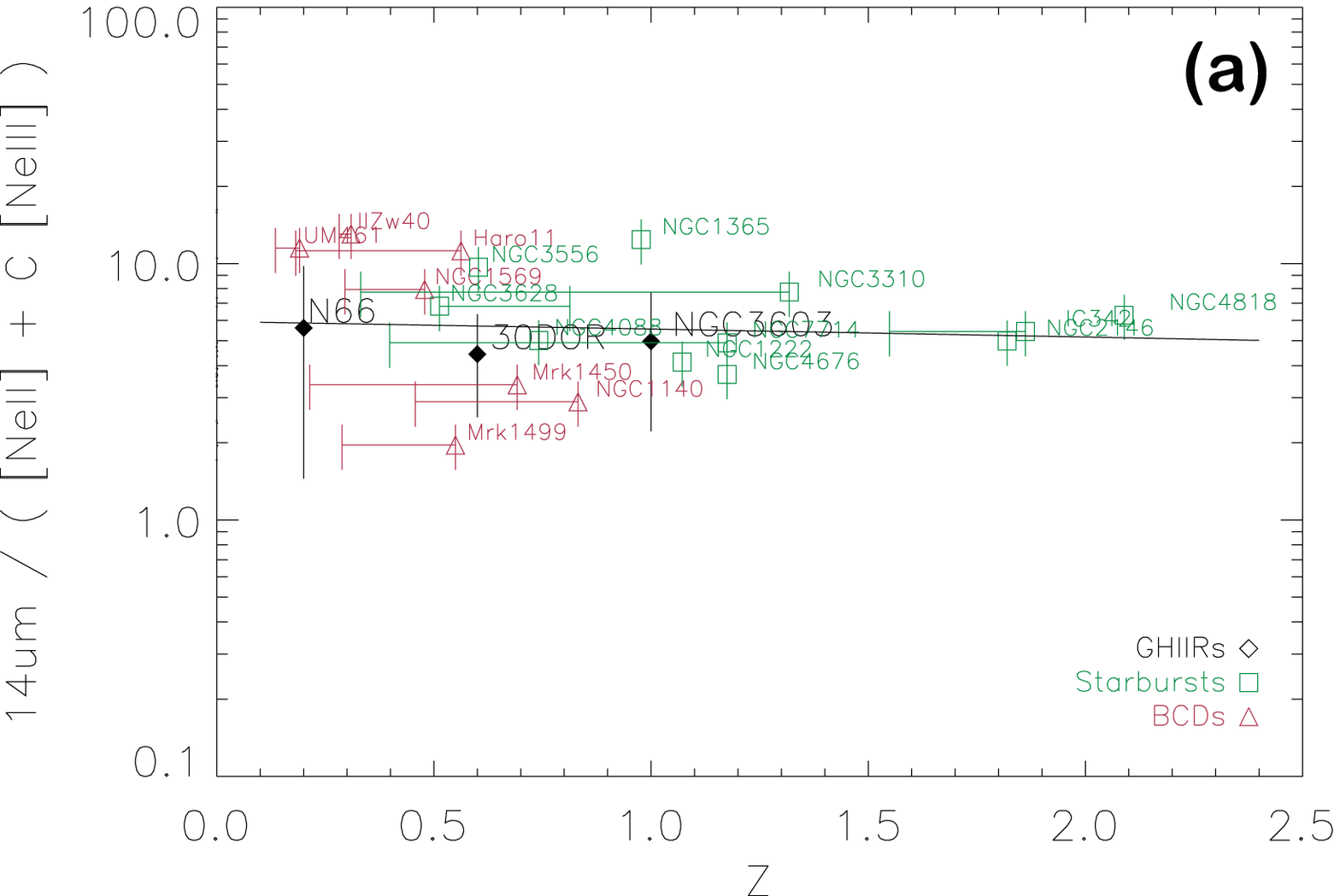}\\
\includegraphics[angle=0,scale=0.5,clip=true]{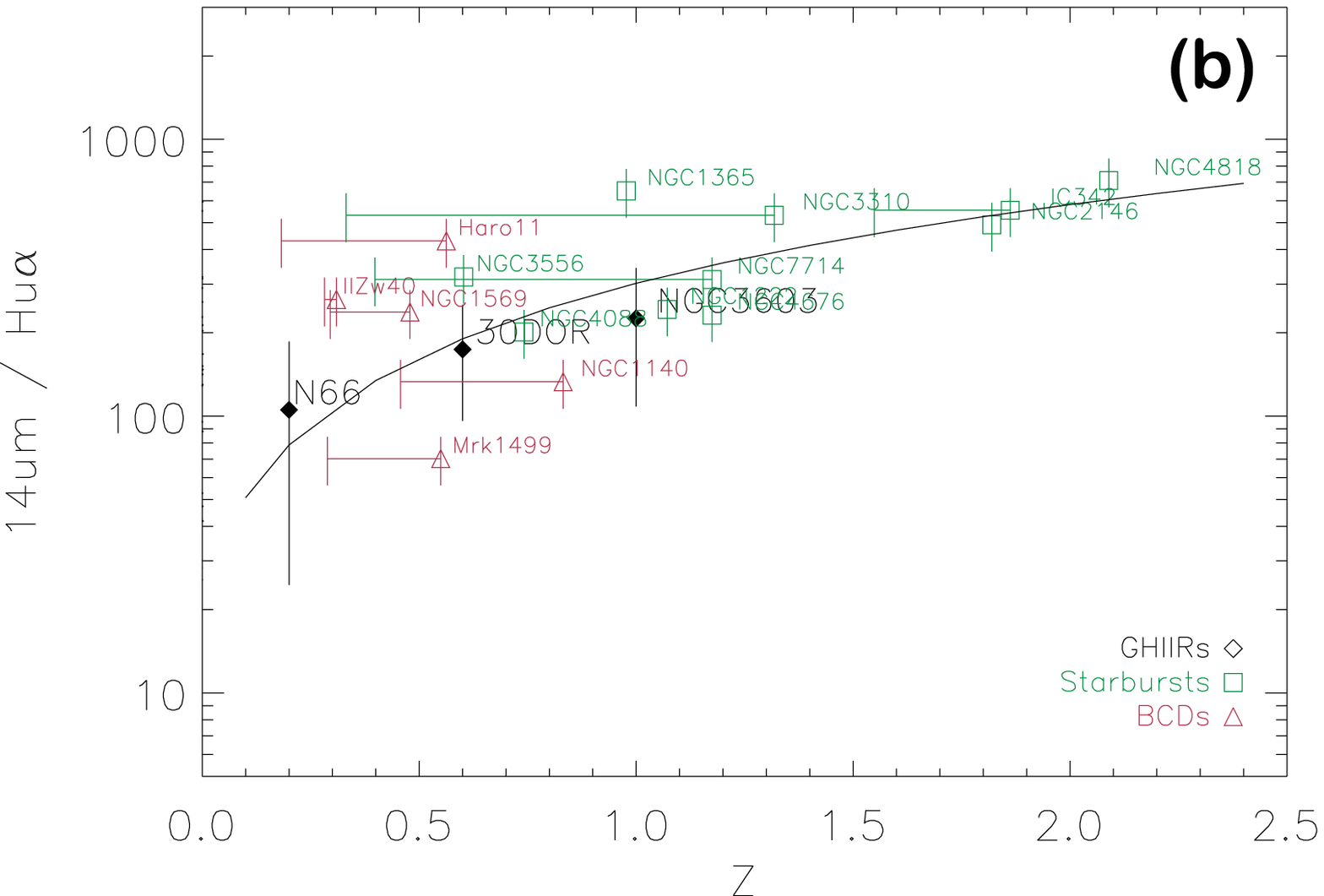}
\figcaption{The 14\mic\ dust grain emission is compared to [Ne\2]+$C$[Ne\3] \textbf{(a)} and Hu$\alpha$ \textbf{(b)} as a function of the metallicity. For clarity, we only display the average ratio in each giant H\2\ region, with the associated standard deviation. For the starburst galaxies and BCDs, we display the metallicity (in solar units) derived from MIR lines with a symbol, and the difference to the optical determination (when it exists) is shown with a horizontal bar. The black solid line shows the polynomial fit to all the data points. \label{fig:vsg_ig_hi}}
\end{figure}

Figure\,\ref{fig:vsg_ig_hi} shows the ratio between the 14\mic\ continuum and ionized gas tracers as a function of metallicity (Sect.\,\ref{sec:obsparam_met}). Two ionized gas tracers are considered, [Ne\2]+$C$ [Ne\3] and the H\1\ recombination line Hu$\alpha$, with the latter having the advantage of not scaling directly with the abundance of heavy elements (Sect.\,\ref{sec:obsparam_ig}). We assume in the following that most of the 14\mic\ dust emission originates in the ionized gas (Sect.\,\ref{sec:obsparam_moldust}). 

Contrary to PAH/14\mic\ or PAH/Hu$\alpha$, the ratio between the 14\mic\ continuum and ionized gas tracers is fairly insensitive to the ISRF hardness. Considering the combined results of the giant H\2\ regions and the star-forming galaxies, the 14\mic/([Ne\2]+$C$ [Ne\3]) ratio does not show any significant correlation with metallicity (Fig.\,\ref{fig:vsg_ig_hi}a). 

This contrasts with the 14\mic/Hu$\alpha$ ratio, especially for low-metallicity regions (Fig.\,\ref{fig:vsg_ig_hi}b). We conclude that the rather constant observed 14\mic/([Ne\2]+$C$ [Ne\3]) ratio is due to the fact that the dust continuum scales with metallicity. Unfortunately, the large dispersion hampers a quantitative determination of the metallicity-dependence. 

Our results showed that the PAH/14\mic\ ratio in N\,66 is relatively higher when compared to positions with similar ISRF hardness in other H\2\ regions (Sect.\,\ref{sec:destr_ghiir}). This discrepancy is consistent with the warm dust continuum emission being reduced in this object because of the relatively low-metallicity environment. The dependence can be explained by a lower dust mass at low-metallicity (see Engelbracht et al.\ 2008).

\section{PAHs vs.\ molecular hydrogen}\label{sec:molh}

Emission from PAH molecules and molecular hydrogen both originates close to or in PDRs (Sect\,\ref{sec:destr}). Since they are carbon-based molecules, the total abundance of PAHs in a galaxy is expected to depend at least partly on the metallicity. On the other hand, H$_2$ formation on dust grains also implies an indirect metallicity-dependence. PAH and H$_2$ survival in harsh radiation field bears important differences since photodissociation of H$_2$ requires energies $\gtrsim$11.2\,eV, while much higher energies are needed for PAHs which are able to thermalize rapidly due to efficient electron delocalization (Sect.\,\ref{sec:destr}).

H$_2$ transitions in the MIR give access to the bulk of the warm molecular phase (few $100$\,K). The most prominent MIR lines are due to the spontaneous deexcitation toward lower rotational levels within the ground vibrational-electronic state. Excitation of H$_2$ can occur in several ways, among which the most frequent are pumping of non-ionizing photons followed by fluorescence, collisional excitation of the lower rotational levels for large densities (few $10^3$\,cm$^{-3}$),  thermal excitation via collisions with photoelectrons, and heating by shocks (see e.g., Roussel et al.\ 2007). 

In UV-dominated regions, PAHs and H$_2$ are expected to trace each other at sufficiently large spatial-scales. Roussel et al.\ (2007) studied H\2\ nuclei within the SINGS sample (Kennicutt et al.\ 2003) and found a tight correlation between PAH emission (bands within the \textit{Spitzer/IRAC} [8\mic]band) and H$_2$ emission (S(0)+ S(1)+S(2) lines). The PAH/H$_2$ ratio has been found by these authors to be fairly insensitive to the radiation field intensity, which reinforces the hypothesis of a similar origin in PDRs. Although with a smaller range of radiation field intensity, Bernard-Salas et al.\ (2009) obtained the same result in their sample of starburst galaxies.

\begin{figure}
\includegraphics[angle=0,scale=0.5,clip=true]{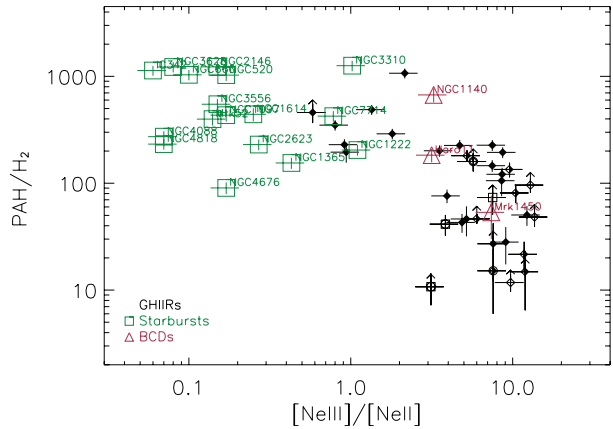}
\figcaption{The PAH emission is compared to H$_2$ as a function of the ISRF hardness. See Fig.\,\ref{fig:isrf2} for the symbol explanation. \label{fig:pah_h2}}
\end{figure}

In order to trace the PAH emission in the objects of our sample, we use the most prominent bands (6.2\mic, 7.7\mic, 8.6\mic, and 11.3\mic), while we use the S(1) and S(2) lines to trace H$_2$ (Sect.~\ref{sec:obsparam_moldust}). Figure~\ref{fig:pah_h2} shows the resulting PAH/H$_2$ ratio as a function of the ISRF hardness. On first order, the PAH and H$_2$ emission scale together if one considers that most of the positions with low PAH/H$_2$ ratios are in fact lower limits. Only a small fraction of the dispersion (which is about a few tens) should be due to the lack of detection of the S(0) line, contributing about 25\% to the total of S(0-2) in H\2\ nuclei (Sect.\,\ref{sec:obsparam_moldust}; Roussel et al.\ 2007). 
A more detailed analysis reveals that the values in NGC\,3603 and 30\,Dor lie at the lower end of the range observed for the starburst galaxies, while the BCD Mark\,1450 and most positions in N\,66 are characterized by significantly lower PAH/H$_2$ ratios. 

The PAH/H$_2$ ratio shows a factor 10 variation among the starburst galaxies. This large dispersion can be attributed to contribution of older stellar populatons to the PAH emission, agreeing with results in Sect.\,\ref{sec:isrf_sfgig}. 

We notice that low PAH/H$_2$ values are only found in objects with hard ISRF, reminding the results obtained in Sect.\,\ref{sec:destr_ghiir}. Hence, it seems that PAH molecules are destroyed in these objects while H$_2$ is still able to emit. Such a result is consistent with H$_2$ being self-shielded so that it is able to emit regardless of the UV field intensity or hardness, contrary to PAH emission which is highly dependent on the physical conditions in the ionized gas. Note that this result supposes that the spectral extraction window (Sect.~\ref{sec:obs_ghiir}) is large enough to include the contribution of both PAH-dominated and H$_2$ dominated regions.

The weak correlation of PAH/H$_2$ with the ISRF hardness is less obvious than what we observed by using PAH/14\mic\ or PAH/Hu$\alpha$ since for a hardness above the threshold for PAH destruction (Sect.~\ref{sec:destr_ghiir}), the PAH/H$_2$ ratio appears to be quite unrelated to the ISRF. Therefore, although these results strongly suggest that PAH and H$_2$ molecules depend differently on the incident UV field, another parameter might explain the range of PAH/H$_2$ for similarly hard ISRF values.

\section{Summary and conclusions}
 
In this paper, we first investigated the PAH emission properties of several sources within three giant H\2\ regions, NGC\,3603 in the Milky Way, 30\,Dor in the LMC, and N\,66 in the SMC. The sample spans an interesting range of metallicities (down to $0.2$\,Z$_\odot$) and physical conditions (ISRF hardness). The sample of giant H\2\ regions contain several ISM components contributing to the MIR spectra of star-forming galaxies, stellar clusters, individual PDRs, warm photoionized gas, as well as embedded MIR bright regions. The main results are:
\begin{itemize}
\item The PAH/14\mic\ ratio shows a strong dependence on the ISRF hardness as traced by [Ne\3]/[Ne\2]. The locations of the data points in the PAH/14\mic\ vs.\ [Ne\3]/[Ne\2] diagram agree well with the PDR vs.\ ionized gas nature of the sources. For soft ISRF, PAH/14\mic\ is uniform and dominated by PDRs. A threshold occurs around [Ne\3]/[Ne\2]$\approx3$ above which the dispersion of PAH/14\mic\ values greatly increases, with a decreasing ratio as the ISRF hardens.  
\item N\,66 is characterized by large PAH/14\mic\ values compared to positions in 30\,Dor with similar physical conditions. The difference is due to suppressed warm dust emission in N\,66, which we propose is an effect of the low dust-to-gas ratio in the ionized gas.
\item The variations of the PAH/14\mic\ ratio are not dominated by varying ISRF conditions leading to different heating mechanism of PAHs and of the dust. A change in the PAH abundance is required, being consistent with the paradigm of photodestruction of PAH molecules in the ionized gas. 
\item PAH emission is compared to the H\1\ recombination line Hu$\alpha$ which traces the ionized gas. The variations of PAH/Hu$\alpha$ are directly related to the relative contribution of PDR vs.\ ionized gas. We find that all sources in the giant H\2\ regions follow the same trend as a function of the ISRF hardness. We use these results as a reference for discussing observations of star-forming galaxies.
\item By investigating the dependence of PAH/Hu$\alpha$ on the iron abundance, we do not find any evidence for enhanced PAH destruction by shocks.
\end{itemize}

The results on the giant H\2\ regions are compared to those of star-forming galaxies (BCDs, starburst galaxies, and H\2\ galaxies), with the following findings:
\begin{itemize}
\item The PAH/14\mic\ ratio  in star-forming galaxies follows on first order the trend of the giant H\2\ regions. Starburst galaxies globally lie with the PDRs in giant H\2\ regions while BCDs lie with sources in giant H\2\ regions dominated by ionized gas.
\item PAH/Hu$\alpha$ in starburst galaxies is larger than PDR values in giant H\2\ regions which is likely due to the PAH excitation by old stellar populations in the diffuse ISM. The PAH emission component in the galaxies of the sample is thus not dominated by a collection of PDRs similar to the ones observed in giant H\2\ regions.
\item We find a metallicity dependence of the 14\mic/Hu$\alpha$ ratio, which reflects the warm dust abundance in the photoionized gas.
\end{itemize}

Low-metallicity star-forming galaxies are particularly challenging environments for detecting PAH emission because of the low carbon abundance which limits the formation of hydrocarbon carriers, and of the hard and intense radiation field which is able to destroy PAH molecules. Furthermore, dwarf galaxies do not contain significant amounts of GMCs, implying little integrated volume of PDRs. Our findings suggest that metal-poor star-forming galaxies such as BCDs have PAH/Hu$\alpha$ ratios that agree on first order with what is expected from typical physical conditions similar to the ionized gas in H\2\ regions. Even though PAHs could be formed in large amounts over the galaxy history, molecules are mostly photodissociated in the ionized gas. In more metal-rich galaxies such as starburst galaxies, there is a sign of significant diffuse PAH emission which could severely bias star-formation rate indicators based on PAH emission. 

\begin{acknowledgements}
The authors acknowledge Dan Weedman, Gregory Sloan, Jim Houck, and Henrik Spoon at Cornell University for their help on this paper. We are also grateful to Yanling Wu who communicated to us some measurements used in this study. VC would like to acknowledge partial support from the EU ToK grant 39965 and FP7-REGPOT 206469.
\end{acknowledgements}

\end{document}